\documentclass[]{iopart}
\usepackage{iopams,graphicx,hyperref}
\usepackage[usenames,dvipsnames]{color}
\usepackage{mathrsfs}
\DeclareSymbolFontAlphabet{\mathrsfs}{rsfs}

\newcommand{\tfrac}[2]{\textstyle \frac{#1}{#2}}
\newcommand{\half}{\tfrac{1}{2}}
\newcommand{\third}{\tfrac{1}{3}}
\newcommand{\fourth}{\tfrac{1}{4}}
\newcommand{\sixth}{\tfrac{1}{6}}

\newcommand{\four}{\,{}^{(4)}}
\newcommand{\fgamma}{\four{\gamma}}
\newcommand{\fg}{\four{g}}
\newcommand{\fR}{\four{R}}
\newcommand{\tR}{\tilde R}
\newcommand{\ftR}{\four{\tilde R}}
\newcommand{\tphi}{\tilde \phi}

\newcommand{\tpsi}{\tilde \psi}
\newcommand{\tBox}{\tilde \Box}
\newcommand{\tGamma}{\tilde \Gamma}
\newcommand{\tnabla}{\tilde \nabla}
\newcommand{\fnabla}{\four{\nabla}}
\newcommand{\ftnabla}{\four{\tilde \nabla}}
\newcommand{\tT}{\tilde T}
\newcommand{\tn}{\tilde n}
\newcommand{\tN}{\tilde N}
\newcommand{\tA}{\tilde A}
\newcommand{\tF}{\tilde F}
\newcommand{\tcD}{\tilde \mathcal{D}}
\newcommand{\tB}{\tilde B}
\newcommand{\tcB}{\tilde \mathcal{B}}
\newcommand{\Lie}{\mathcal{L}}
\newcommand{\const}{\mathrm{const}}
\newcommand{\scri}{$\mathrsfs{I}^+\,$}
\newcommand{\hateq}{\; \hat{=} \;}
\renewcommand{\Or}{\mathcal{O}}
\newcommand{\rmin}{r_\mathrm{min}}
\newcommand{\rAH}{r_\mathrm{AH}}
\newcommand{\MB}{M_\mathrm{B}}
\newcommand{\MAH}{M_\mathrm{AH}}
\newcommand{\MBi}{M_\mathrm{B}^\mathrm{i}}
\newcommand{\MAHi}{M_\mathrm{AH}^\mathrm{i}}
\newcommand{\MBf}{M_\mathrm{B}^\mathrm{f}}
\newcommand{\MAHf}{M_\mathrm{AH}^\mathrm{f}}

\begin{document}

\title{Hyperboloidal Einstein-matter evolution and
  tails for scalar and Yang-Mills fields}

\author{Oliver Rinne$^{1}$ and Vincent Moncrief$^{2}$}
\address{$^{1}$Max Planck Institute for Gravitational Physics 
  (Albert Einstein Institute), Am M\"uhlenberg 1, 14476 Potsdam, Germany\\
  $^{2}$Department of Mathematics and Department of Physics, Yale University, 
  New Haven, CT 06520, USA}
\ead{oliver.rinne@aei.mpg.de, vincent.moncrief@yale.edu}
\date{\today}

\begin{abstract}
  We show how matter can be included in a constrained ADM-like formulation
  of the Einstein equations on constant mean curvature surfaces.
  Previous results on the regularity of the equations at future null infinity
  are unaffected by the addition of matter with tracefree energy-momentum
  tensor.
  Two examples are studied in detail, a conformally coupled scalar field
  and a Yang-Mills field.
  We first derive the equations under no symmetry assumptions and then reduce
  them to spherical symmetry.
  Both sectors (gravitational and sphaleron) of the spherically symmetric 
  Yang-Mills field are included.
  We implement this scheme numerically in order to study late-time tails
  of scalar and Yang-Mills fields coupled to the Einstein equations.
  We are able to evolve spacetimes that disperse to flat space,
  accrete onto a given black hole or collapse to a black hole from
  regular initial data.
  The sphaleron sector of Yang-Mills is found to exhibit some nontrivial
  gauge dynamics.
\end{abstract}

%%%%%%%%%%%%%%%%%%%%%%%%%%%%%%%%%%%%%%%%%%%%%%%%%%%%%%%%%%%%%%%%%%%%%%%%%%%%%%%
%%%%%%%%%%%%%%%%%%%%%%%%%%%%%%%%%%%%%%%%%%%%%%%%%%%%%%%%%%%%%%%%%%%%%%%%%%%%%%%

\section{Introduction}
\label{s:intro}

Most current numerical relativity codes are based on the Cauchy formulation of 
general relativity and evolve spacetime on spacelike slices
approaching spacelike infinity, truncated at some finite distance.
At the resulting artificial timelike boundary, boundary conditions must be
imposed that yield a well-posed initial-boundary value problem and are 
compatible with the constraint equations. 
In addition, for evolutions of isolated systems, gravitational radiation
should pass through the boundary without causing spurious reflections, 
i.e.~the boundary conditions should be \emph{absorbing}.
Considerable progress has been made recently with the construction and 
implementation of boundary conditions for the Einstein equations 
(see \cite{SarbachLRR} for a recent review article).
A fundamental problem remains, however.
In general relativity there is no well-defined flux of gravitational radiation
at a finite distance upon which absorbing boundary conditions could be based.
At best one may appeal to linearised theory.
Gravitational radiation is only well defined at future null infinity \scri.
Thus a far more elegant solution to the outer boundary problem is to
include \scri in the numerical domain.

We follow Penrose's approach \cite{Penrose1965} and apply a conformal 
transformation to the spacetime metric, combined with a compactifying
coordinate transformation that maps an asymptotically flat spacetime
to a finite domain.
In \cite{Moncrief2009} we developed an ADM-like \cite{Arnowitt1962} formulation of the
vacuum Einstein equations on constant mean curvature (CMC) surfaces.
These are spacelike but approach future null infinity instead of spacelike
infinity; thus one solves a \emph{hyperboloidal} initial value problem.
(Note that hyperboloidal surfaces are not Cauchy surfaces; we only
obtain the part of spacetime to the future of the initial hyperboloidal 
surface.)
Since the Ricci tensor is not conformally invariant, the Einstein equations
contain inverse powers of the conformal factor that are singular at \scri.
However, in \cite{Moncrief2009} we showed how the formally singular terms in the ADM
evolution equations can be evaluated at \scri in a regular way provided
the constraints hold and \scri is shear free.
In \cite{Rinne2010} this scheme was implemented numerically for axisymmetric
spacetimes.
Long-term stable evolutions of a gravitationally perturbed Schwarzschild black 
hole were obtained and the Bondi news function describing the gravitational 
radiation emitted by the system was evaluated at \scri.

Before continuing we briefly review other hyperboloidal evolution schemes
and associated numerical studies.
The oldest and, arguably, mathematically best understood formulation are the 
regular conformal field equations due to Friedrich \cite{Friedrich1983a}.
These form a symmetric hyperbolic system of partial differential equations that
contain the Einstein equations as well as evolution equations for the Weyl
curvature arising from the Bianchi identities.
The equations have the remarkable property that they are manifestly regular 
up to \scri.
There have been various attempts at numerical evolutions based on these
equations (see the review articles \cite{FrauendienerLRR,Husa2002,Husa2003}).
With a view to the applications considered in the present paper, we mention in
particular the studies of spherically symmetric scalar field collapse by 
H\"ubner \cite{Huebner1995,Huebner1996}.
One difference to our setup is that in these studies, the generators 
of \scri converge and future timelike infinity $i^+$ is reached in finite 
computational time, whereas we hold the coordinate location of \scri fixed 
and only approach $i^+$ asymptotically in the limit of infinite computational
time. 
The latter appears to be better suited for studying radiative phenomena as it
does not suffer from a loss of numerical resolution as $i^+$ is approached.

Recently a number of formulations have been suggested that are based more
directly on the Einstein equations, as in our approach.
Zengino\u{g}lu \cite{Zenginoglu2008} developed a formulation based on 
generalised harmonic gauge combined with a suitable choice of gauge source 
functions at \scri.
Bardeen, Sarbach and Buchman \cite{Bardeen2011} derived a tetrad formulation
of the Einstein equations on CMC slices.
First numerical results on initial data for single and binary 
black holes were presented in \cite{Buchman2009,Bardeen2012}.

In this paper we return to the constrained ADM formulation on CMC slices 
developed in \cite{Moncrief2009} and extend it to include matter sources.
The motivation for this derives partly from the fact that we wanted to study
the late-time behaviour of perturbed black hole spacetimes in the context of
the full (rather than linearised) Einstein equations, including future null 
infinity.
In \cite{Rinne2010} we correctly reproduced the quasi-normal mode radiation
emitted by a perturbed black hole but due to limited numerical resolution we
were unable to resolve the power-law tail expected at later times.
Therefore, in order to see if our method is suitable to study these phenomena,
we decided to take one step back and consider spherically symmetric spacetimes,
which are computationally less expensive to evolve.
Because of Birkhoff's theorem, matter is needed in order to have nontrivial 
dynamics in spherical symmetry.
How to include matter in a hyperboloidal Einstein evolution scheme is an
interesting problem in its own right.

At late times matter fields as well as gravitational perturbations on
flat space and black hole spacetimes typically decay polynomially in time,
a phenomenon often referred to as Price's law \cite{Price1972}.
This power-law tail is caused by the backscatter off the curved background
spacetime and/or by the nonlinearity of the matter fields.
It plays an important role in trying to prove stability of black hole
spacetimes and the cosmic censorship 
conjecture \cite{Dafermos2005}.
At \scri the fields generally decay at a slower rate than at any finite 
distance, although the closer an observer is to \scri, the longer the measured
decay rate stays close to the value corresponding to \scri before it ultimately
approaches the faster finite-distance decay.
It can be argued \cite{Puerrer2005} that the decay rate at \scri is the relevant
one for observers in the astrophysical zone \cite{Leaver1986}, in which the 
distance to the source of radiation is very large compared to the time during
which the signal is observable.

For massless scalar fields the tail decay rates can be predicted from linear 
perturbation theory \cite{Gundlach1994,Gundlach1994a} and have been confirmed 
numerically many times.
We mention two recent studies that both include \scri in the numerical domain.
P\"urrer, Husa and Aichelburg \cite{Puerrer2005} evolved the spherically 
symmetric Einstein-scalar field system in Bondi coordinates and determined 
the power-law tails in subcritical evolutions that disperse to flat space.
Bondi coordinates cannot penetrate horizons so the decay of the field at the 
horizon of a black hole could not be studied.
Zengino\u{g}lu \cite{Zenginoglu2008b} evolved the spherically symmetric scalar 
wave equation on a fixed background spacetime (taken to be either Minkowski or 
Schwarzschild), the test field approximation.
He used a hyperboloidal foliation of spacetime that covers part of the black 
hole interior as well.
In the present paper we evolve the coupled Einstein-scalar field equations on
hyperboloidal slices reaching out to \scri.
In particular, we are able to study the decay of the field at the horizon
of a black hole formed in gravitational collapse.

For Yang-Mills fields the prediction from linear perturbation theory turns out
to be incorrect: the nonlinearity of the field causes a slower 
decay \cite{Bizon2007}.
Similarly to the scalar field, the Einstein-Yang-Mills system was evolved
in Bondi coordinates in \cite{Puerrer2009} and in the test field approximation 
on hyperboloidal slices in \cite{Zenginoglu2008b}.
We shall compare our results with those studies.

This paper is organised as follows.
In section \ref{s:general} we extend our hyperboloidal Einstein evolution 
scheme \cite{Moncrief2009} to include matter sources
and we re-examine the question of regularity at \scri.
Two matter models are studied in detail in section \ref{s:matter},
a conformally coupled scalar field and a Yang-Mills field.
In section \ref{s:spher} we reduce our formulation to spherical symmetry.
The numerical implementation of this system is described in 
section \ref{s:nummethod} and our results on power-law tails are presented in
section \ref{s:numresults}.
Finally we conclude and discuss some directions for future work in
section \ref{s:concl}.
Some useful identities for conformal transformations and $3+1$ decompositions 
are collected in \ref{s:app}.

%%%%%%%%%%%%%%%%%%%%%%%%%%%%%%%%%%%%%%%%%%%%%%%%%%%%%%%%%%%%%%%%%%%%%%%%%%%%%%%
%%%%%%%%%%%%%%%%%%%%%%%%%%%%%%%%%%%%%%%%%%%%%%%%%%%%%%%%%%%%%%%%%%%%%%%%%%%%%%%

\section{General formalism}
\label{s:general}

In this section we consider general matter subject to the condition that its
energy-momentum tensor be tracefree.
We extend the conformal ADM formulation of the Einstein equations derived 
in \cite{Moncrief2009} to include the corresponding matter source terms.
The question of regularity at future null infinity is re-examined.

\subsection{Matter in the conformal setting}

We use the notation and conventions of \cite{Moncrief2009}.
The spacetime metric $\four g_{\mu\nu}$ is written as
\begin{equation}
  \label{e:confmetric}
  \four g_{\mu\nu} = \Omega^{-2} \four \gamma_{\mu\nu},
\end{equation}
where $\four \gamma_{\mu\nu}$ is the conformal spacetime metric and $\Omega$
the conformal factor.
Consider now matter given by an energy-momentum tensor $T_{\mu\nu}$.
We introduce a conformally rescaled energy-momentum tensor $\tT_{\mu\nu}$ via
\begin{equation}
  \label{e:tT}
  T_{\mu\nu} = \Omega^2 \tT_{\mu\nu}.
\end{equation}
The energy-momentum conservation equations transform as \cite{Friedrich1991}
\begin{equation}
  \label{e:emtrafo}
  \fgamma^{\mu\nu} \ftnabla_\mu \tT_{\nu \rho} = \Omega^{-4} \fg^{\mu\nu} \left(
    \fnabla_\mu T_{\nu\rho} - \Omega^{-1} \fnabla_\rho \Omega \, T_{\mu\nu} \right),
\end{equation}
where $\fnabla$ denotes the covariant derivative of 
$\four g$ and $\ftnabla$ the covariant derivative of $\four\gamma$.
The first term on the right-hand side of \eref{e:emtrafo} vanishes by 
energy-momentum conservation.
If the energy-momentum tensor is tracefree,
\begin{equation}
  \label{e:Ttf}
  \fg^{\mu\nu} T_{\mu\nu}=0,
\end{equation} 
then the second term on the right-hand side of \eref{e:emtrafo} also vanishes 
and the energy-momentum conservation equations reduce to
\begin{equation}
  \label{e:emtrafo1}
  \fgamma^{\mu\nu} \ftnabla_\mu \tT_{\nu \rho} = 0.
\end{equation}
Assuming that $\tT_{\mu \nu}$ is itself regular at \scri, which it is if its
basic fields satisfy conformally regular field equations (as in the matter
models considered in section \ref{s:matter}), then \eref{e:emtrafo1} is
manifestly regular at future null infinity \scri.
For this reason we will restrict ourselves to matter with tracefree 
energy-momentum tensor.

Examples of matter models satisfying this condition include the conformally 
coupled scalar field (section \ref{s:ccscalar}), Maxwell and 
more generally Yang-Mills fields (section \ref{s:ym}) and the radiation fluid,
i.e.~a perfect fluid with equation of state $p=\third\rho$ 
(see also \cite{Luebbe2011a}).
There are of course many matter models that do not satisfy \eref{e:Ttf}.
Typically however, these are considered to be less radiative.
For example, a massive scalar field is known to fall off faster than any power
of radius towards \scri \cite{Winicour1988}.
If the matter remains bounded away from \scri, as is expected in many situations
of astrophysical interest, then of course there is no harm in using the 
singular matter field equations away from \scri.

%%%%%%%%%%%%%%%%%%%%%%%%%%%%%%%%%%%%%%%%%%%%%%%%%%%%%%%%%%%%%%%%%%%%%%%%%%%%%%%

\subsection{Conformal ADM reduction of the Einstein equations with 
  matter sources}
\label{s:generalsources}

As in \cite{Moncrief2009} we decompose the physical and conformal spacetime 
metric in $3+1$ form,
\begin{eqnarray}
  \four g = -N^2 \rmd t^2 + g_{ij} (\rmd x^i + X^i \rmd t)(\rmd x^j + X^j \rmd t)
  ,\\
  \four \gamma = -\tN^2 \rmd t^2 + \gamma_{ij} (\rmd x^i + X^i \rmd t)
  (\rmd x^j + X^j \rmd t),
\end{eqnarray}
where the physical and conformal lapse are related via $N = \Omega^{-1} \tN$,
the physical and conformal spatial metric via $g = \Omega^{-2} \gamma$,
and $X^i$ serves both as physical and conformal shift.
The unit timelike normals to the $t=\const$ hypersurfaces in physical and 
conformal spacetime are related by $n^\mu = \Omega \tn^\mu$.

The extrinsic curvature is defined as
\begin{equation}
  K_{ij} = -\half \Lie_n g_{ij} . 
\end{equation}  
Our condition on the time slices is that their mean extrinsic curvature be 
constant,
\begin{equation}
  \label{e:constk}
  g^{ij} K_{ij} \equiv -K = \const 
\end{equation}  
with $K>0$ so that the slices approach future null infinity (hence our slightly
awkward sign convention in \eref{e:constk}).
We choose to work with the traceless part of the ADM momentum, which is given
in terms of the extrinsic curvature by
\begin{equation}
  \pi^{\tr ij} = -\mu_g (g^{ik} g^{jl} - \third g^{ij} g^{kl}) K_{kl},
\end{equation}
where $\mu_g = \sqrt{\det(g_{ij})}$.
For later use we also define the conformal extrinsic curvature
\begin{equation}
  C_{ij} = -\half \Lie_{\tn} \gamma_{ij} . 
\end{equation}  
Its trace $C \equiv \gamma^{ij} C_{ij}$ is related to the variable 
$\Gamma = -2\tN C$ introduced in \cite{Moncrief2009}.

Following \cite{York1979} we define the following projections of the 
energy-momentum tensor,
\begin{equation}
  \rho \equiv n^\mu n^\nu T_{\mu\nu}, \qquad
  J^i \equiv -g^{i\mu}n^\nu T_{\mu\nu}, \qquad
  S_{ij} \equiv g_i{}^\mu g_j{}^\nu T_{\mu\nu},
\end{equation}
and similarly for the conformally rescaled energy-momentum tensor \eref{e:tT},
\begin{equation}
  \label{e:cTTprojs}
  \tilde \rho \equiv \tn^\mu \tn^\nu \tT_{\mu\nu}, \qquad
  \tilde J^i \equiv -\gamma^{i\mu}\tn^\nu \tT_{\mu\nu}, \qquad
  \tilde S_{ij} \equiv \gamma_i{}^\mu \gamma_j{}^\nu \tT_{\mu\nu}.
\end{equation}
Clearly these quantities are related by
\begin{equation}
  \rho = \Omega^4 \tilde \rho, \qquad
  J^i = \Omega^5 \tilde J^i, \qquad
  S_{ij} = \Omega^2 \tilde S_{ij}.
\end{equation}

The Einstein equations are
\begin{equation}
  G_{\mu\nu} = \kappa T_{\mu\nu}
\end{equation}
with $\kappa = 8\pi$ in geometric units (i.e.~Newton's constant and the
speed of light $G = c = 1$).
They split into evolution equations and constraints. 
The evolution equations are (cf.~equations (2.10), (2.21) and (2.22) 
in \cite{Moncrief2009})
\begin{eqnarray}
  \label{e:dtOmega}
  \fl \Lie_{\tn} \Omega = -\third (K + \Omega C),\\
  \label{e:dtgamma}
  \fl \Lie_{\tn} \gamma_{ij} = 2 \mu_\gamma^{-1} \gamma_{ik}\gamma_{jl} \pi^{\tr kl}
     - \tfrac{2}{3} \gamma_{ij} C,\\
  \label{e:dtpi}
  \fl \Lie_{\tn} \pi^{\tr ij} = 
  - 2 \mu_\gamma^{-1} \gamma_{kl} \pi^{\tr ik} \pi^{\tr jl}
  - \tfrac{2}{3} \Omega^{-1} K \pi^{\tr ij}\nonumber\\
  + \mu_\gamma \left[ \tN^{-1} \tnabla^i\tnabla^j \tN -  \tR^{ij}
  - 2  \Omega^{-1} \tnabla^i\tnabla^j\Omega
  + \kappa \Omega^2 \tilde S^{ij} \right]^{\tr}.
\end{eqnarray}
Here and in the following, indices on quantities carrying a tilde are to be 
raised and lowered with $\gamma$, $\tr$ denotes the traceless part 
w.r.t.~$\gamma$, $\tR_{ij}$ is the Ricci tensor of $\gamma$, and $\Lie$ 
denotes the Lie derivative.
Note that $\pi^{\tr ij}$ is a tensor density of weight one and hence
\begin{equation}
  \Lie_{\tn} \pi^{\tr ij} = \tN^{-1} \left[ \partial_t \pi^{\tr ij} 
  - (X^k \pi^{\tr ij})_{,k} + X^i{}_{,k} \pi^{\tr kj} + X^j{}_{,k} \pi^{\tr ik} \right].
\end{equation}
The Hamiltonian and momentum constraints are (cf.~equations (2.9) and (2.7) 
in \cite{Moncrief2009})
\begin{eqnarray}
  \label{e:hamcons}
  \fl 0 = -4\Omega \tnabla^j\tnabla_j\Omega + 6 \gamma^{ij}\Omega_{,i}\Omega_{,j}
  - \Omega^2\tR -\tfrac{2}{3} K^2 + \Omega^2 \mu_\gamma^{-2} \gamma_{ik}\gamma_{jl} 
  \pi^{\tr ij}\pi^{\tr kl} + 2 \kappa \Omega^4 \tilde \rho,\\
  \label{e:momcons}
  \fl 0 = \tnabla_j(\Omega^{-2}\pi^{\tr ij}) 
  + \kappa \mu_\gamma \tilde J^i.
\end{eqnarray}

Preservation of the CMC condition, $\partial_t K = 0$, implies an elliptic 
equation for the conformal lapse (cf.~equation (2.13) in \cite{Moncrief2009}),
\begin{eqnarray}
  \label{e:cmc}
  \fl -\Omega^2\gamma^{ij}\tnabla_i\tnabla_j\tN 
  + 3\Omega\gamma^{ij}\tN_{,i}\Omega_{,j} 
  -\tfrac{3}{2} \tN \gamma^{ij}\Omega_{,i}\Omega_{,j} + \sixth \tN K^2
    -\fourth \tN\Omega^2\tR \nonumber\\
   + \tfrac{5}{4}\tN \Omega^2 \mu_\gamma^{-2}\gamma_{ik}\gamma_{jl} 
     \pi^{\tr ij}\pi^{\tr kl}
    + \half \kappa \tN \Omega^4 (\tilde S + 2 \tilde \rho) = 0,
\end{eqnarray}
where $\tilde S \equiv \gamma^{ij}\tilde S_{ij}$.

We also need to specify the spatial coordinates.
In \cite{Moncrief2009} we imposed a spatially harmonic gauge condition, 
which yielded an elliptic system for the shift (equation (2.15) in 
\cite{Moncrief2009}).
However, other choices are possible. 
For example, in \cite{Rinne2010} and in section \ref{s:spher} below we use 
coordinates adapted to a spacetime symmetry.

There is a residual conformal gauge freedom inherent in the decomposition
\eref{e:confmetric}.
In \cite{Moncrief2009} we fixed this by requiring the conformal scalar curvature
$\tR$ to be constant; this resulted in an elliptic equation for 
$\Gamma = -2\tN C$ (equation (2.12) in \cite{Moncrief2009}).
In section \ref{s:spher} the spherically symmetric conformal metric will be
taken to be flat; this eliminates the conformal gauge freedom.

%%%%%%%%%%%%%%%%%%%%%%%%%%%%%%%%%%%%%%%%%%%%%%%%%%%%%%%%%%%%%%%%%%%%%%%%%%%%%%%%

\subsection{Regularity at future null infinity}
\label{s:scrireg}

The evolution equation for the traceless momentum \eref{e:dtpi} and the elliptic
equations \eref{e:hamcons}--\eref{e:cmc} contain inverse powers
of the conformal factor that are singular at \scri.
However in \cite{Moncrief2009}  we showed, for the vacuum case, that the formally 
singular terms in \eref{e:dtpi} can in fact be evaluated at \scri in terms of
regular conformal quantities, provided the constraints are satisfied and
\scri is shear free (see also \cite{Andersson1992}).
We also showed that these regularity conditions are preserved under the time 
evolution.

Our analysis exploited the fact that the elliptic equations 
\eref{e:hamcons}--\eref{e:cmc} are degenerate at \scri.
This allows one to evaluate the first few radial derivatives of the fields 
explicitly at \scri, where radius $r$ refers to a coordinate on a given spatial
slice that is constant on the cut of the slice with \scri.
In particular, we obtained expressions for the first three radial derivatives
of $\Omega$, the first two radial derivatives of $\tN$ and the first radial
derivatives of $\pi^{\tr ri}$ at \scri.
This information was sufficient in order to evaluate the formally singular
terms in \eref{e:dtpi} by applying L'Hospital's rule.

It is easy to see that the addition of matter does not affect those results,
essentially because the matter terms in \eref{e:hamcons}--\eref{e:cmc} are 
multiplied by sufficiently high powers of $\Omega$ that vanish at \scri even
after a certain number of derivatives are taken.
More specifically, the third radial derivative of $\Omega$ at \scri was 
obtained in \cite{Moncrief2009} by evaluating the Laplacian of \eref{e:hamcons}.
Now the Laplacian of the matter contribution to \eref{e:hamcons} is 
$\Or(\Omega^2)$ and hence does not contribute at \scri.
Clearly, the matter contributions to lower derivatives of $\Omega$ are 
multiplied by even higher powers of $\Omega$ and vanish at \scri as well.
The first radial derivative of $\pi^{\tr ri}$ was obtained by first multiplying
\eref{e:momcons} by $\Omega^3$ and then taking a radial derivative.
After this operation the matter contribution is $\Or(\Omega^2)$ and hence, 
again, does not contribute at \scri.
We found the second radial derivative of $\tN$ by forming the
linear combination $\Omega^{-1}\times \eref{e:cmc} 
+ \fourth \Omega^{-1} \tN \times \eref{e:hamcons}$ and then taking a radial 
derivative. 
The matter contribution to the resulting expression is $\Or(\Omega^2)$ and hence
does not contribute at \scri.
Thus all the expressions we derived in order to evaluate the formally 
singular terms in \eref{e:dtpi} at \scri are unchanged by the addition of 
matter.
The matter term in \eref{e:dtpi} itself is $\Or(\Omega^2)$ and
thus vanishes at \scri.

%%%%%%%%%%%%%%%%%%%%%%%%%%%%%%%%%%%%%%%%%%%%%%%%%%%%%%%%%%%%%%%%%%%%%%%%%%%%%%%
%%%%%%%%%%%%%%%%%%%%%%%%%%%%%%%%%%%%%%%%%%%%%%%%%%%%%%%%%%%%%%%%%%%%%%%%%%%%%%%

\section{Examples of matter models}
\label{s:matter}

In this section we consider two matter models in detail, a conformally coupled
scalar field and Yang-Mills theory.
We write the matter equation of motion in $3+1$ form and work out the 
projections of the energy-momentum tensor that appear as source terms 
in the $3+1$ form of the Einstein equations.

\subsection{The conformally coupled scalar field}
\label{s:ccscalar}

The action for the Einstein-scalar field model with conformal coupling is 
given by
\begin{equation}
  \label{e:cc_action}
  S = \int \left( \tfrac{1}{2\kappa} \fR 
    - \half \fg^{\mu\nu} \phi_{,\mu} \phi_{,\nu} 
    - \tfrac{1}{12} \fR \phi^2 \right) \mu_{\fg} \, \rmd^4 x.
\end{equation}
This differs from minimal coupling by the additional last term containing the
four-dimensional Ricci scalar. 
As we shall see shortly, the advantage of the conformally coupled model is that
it yields a conformally invariant evolution equation for the scalar field.
We note however that the conformally coupled Einstein-scalar field equations 
are equivalent to the minimally coupled ones in the following 
sense \cite{Bekenstein1974,Huebner1995}:
Suppose $(\phi^\mathrm{cc},\four g_{\mu\nu}^\mathrm{cc})$ is a solution to the 
conformally coupled equations.
Then $(\phi^\mathrm{mc},\four g_{\mu\nu}^\mathrm{mc})$ is a solution to the 
minimally coupled equations, where
\begin{equation}
  \phi^\mathrm{mc} = \sqrt{6}\,\mathrm{arctanh}(\half \phi^\mathrm{cc}), \qquad
  \four g_{\mu\nu}^\mathrm{mc} = [1 - \fourth(\phi^\mathrm{cc})^2] 
  \four g_{\mu\nu}^\mathrm{cc}.
\end{equation}

Varying \eref{e:cc_action} w.r.t. $\phi$ yields the equation of motion
\begin{equation}
  \label{e:cc_eom}
  \Box \phi - \sixth \fR \, \phi = 0.
\end{equation}
Defining a conformally rescaled field $\tphi \equiv \Omega^{-1} \phi$ and using
the transformation of the Ricci scalar \eref{e:Rtrafo}, we find 
\begin{equation}
  \label{e:cc_eom1}
  \tBox \tphi - \sixth \ftR \, \tphi 
   = \Omega^{-3} (\Box \phi - \sixth \fR \, \phi) = 0.
\end{equation}
The left-hand side of this equation is manifestly regular at \scri.

Varying \eref{e:cc_action} w.r.t. $\fg_{\mu\nu}$ produces Einstein's equations
$G_{\mu\nu} = \kappa T_{\mu\nu}$ with energy-momentum tensor
\begin{equation}
  \fl T_{\mu\nu} = \phi_{,\mu}\phi_{,\nu} - \half\phi\fnabla_\mu\fnabla_\nu\phi
  + \fourth\phi^2\fR_{\mu\nu} 
  - \fourth \fg_{\mu\nu}(\fg^{\rho\sigma}\phi_{,\rho}\phi_{,\sigma}
  + \sixth \phi^2 \fR),
\end{equation}
where an overall factor of $\tfrac{2}{3}$ has been absorbed into the
definition of $\phi$, and the equation of motion \eref{e:cc_eom} has been used
to eliminate a $\Box \phi$ term.
This can also be written as $T_{\mu\nu}=\Omega^2 \tT_{\mu\nu}$, where
the conformally rescaled energy-momentum tensor has exactly the same form,
\begin{equation}
  \label{e:cc_emtensor}
  \fl \tT_{\mu\nu} = \tphi_{,\mu}\tphi_{,\nu}
  - \half\tphi\ftnabla_\mu\ftnabla_\nu\tphi
  + \fourth\tphi^2\ftR_{\mu\nu} 
  -\fourth \fgamma_{\mu\nu}(\fgamma^{\rho\sigma}\tphi_{,\rho}\tphi_{,\sigma}
  + \sixth \tphi^2 \ftR).
\end{equation}

Using the $3+1$ decomposition of the scalar field Hessian (\ref{s:scalhess})
and the conformal spacetime Ricci tensor (\ref{s:confricc}),
the equation of motion \eref{e:cc_eom1} can be written as
\begin{eqnarray}
  \label{e:cc_eom2}
  \fl 0 = -\Lie_{\tn}^2 \tphi + \tnabla_i\tnabla^i \tphi 
  + C \Lie_{\tn}\tphi + \tN^{-1} \gamma^{ij} \tN_{,i} \tphi_{,j}\nonumber\\
  - \sixth \tphi \left(
    -2 \tN^{-1}\tnabla^i\tnabla_i\tN + \tR + \tfrac{4}{3} C^2 
    + \mu_\gamma^{-2} \gamma_{ik}\gamma_{jl}\pi^{\tr ij}\pi^{\tr kl} - 2 \Lie_{\tn} C 
  \right)
\end{eqnarray}
(Recall that we are raising and lowering spatial indices with $\gamma_{ij}$.)
In general, the time derivative of the conformal mean curvature $C$ cannot 
be evaluated explicitly during an evolution, e.g. if $\Gamma = -2\tN C$
is determined by solving an elliptic equation as 
in \cite{Moncrief2009}.
This problem can be avoided by introducing a new variable $\tpsi$ and 
writing \eref{e:cc_eom2} as a first-order (in time) system
\begin{eqnarray}
  \label{e:cc_dtphi}
  \fl \Lie_{\tn}\tphi = \tpsi + \third C \tphi,\\
  \label{e:cc_dtpsi}
  \fl \Lie_{\tn}\tpsi = \tnabla_i\tnabla^i \tphi + \tfrac{2}{3} C\tpsi 
  + \tN^{-1} \gamma^{ij} \tN_{,i} \tphi_{,j}\nonumber\\
  - \sixth \tphi \left(
    -2 \tN^{-1}\tnabla^i\tnabla_i\tN + \tR 
    + \mu_\gamma^{-2} \gamma_{ik}\gamma_{jl}\pi^{\tr ij}\pi^{\tr kl} \right).
\end{eqnarray}

The matter source terms in the geometry equations evaluate to
\begin{eqnarray}
  \label{e:cc_rho}
  \fl \tilde \rho = -\half \tphi \tnabla_i \tnabla^i \tphi 
  + \fourth \tphi_{,i} \tphi^{,i}  + \tfrac{3}{4}\tpsi^2
  + \tfrac{1}{8} \tphi^2 \tR 
  -\tfrac{1}{8} \tphi^2 \mu_\gamma^{-2} \gamma_{ik}\gamma_{jl}\pi^{\tr ij}\pi^{\tr kl},
  \\
  \fl \tilde J^i = \gamma^{ij} (-\tphi_{,j}\tpsi + \half \tphi \tpsi_{,j}) 
  - \fourth \mu_\gamma^{-1} \tnabla_j (\tphi^2 \pi^{\tr ij}),
  \\
  \fl \tilde S = \tilde \rho,
  \\
  \label{e:cc_Str}
  \fl \tilde S^{\tr ij} = - \half \tphi [\tnabla^i \tnabla^j \tphi]^{\tr}
  + [\tphi^{,i}\tphi^{,j}]^{\tr} + \half \tphi \tpsi \mu_\gamma^{-1}\pi^{\tr ij}
  \nonumber\\
  \fl \qquad \qquad + \fourth \tphi^2 \left( \mu_\gamma^{-1}\Lie_{\tn}\pi^{\tr ij}
    + 2 \mu_\gamma^{-2}\gamma_{kl}\pi^{\tr ik}\pi^{\tr jl}
    - \tN^{-1} [\tnabla^i \tnabla^j \tN]^{\tr} + \tR^{\tr ij} \right).
\end{eqnarray}
The equation of motion \eref{e:cc_eom2} has been used in order to 
eliminate a $\Lie_{\tn}^2 \tphi$ term in \eref{e:cc_rho}.
Note that \eref{e:cc_Str} contains $\Lie_{\tn}\pi^{\tr ij}$.
When \eref{e:cc_Str} is substituted in the evolution equation 
\eref{e:dtpi} for the traceless momentum, the resulting equation must be 
solved for $\Lie_{\tn}\pi^{\tr ij}$.

%%%%%%%%%%%%%%%%%%%%%%%%%%%%%%%%%%%%%%%%%%%%%%%%%%%%%%%%%%%%%%%%%%%%%%%%%%%%%%%

\subsection{Yang-Mills theory}
\label{s:ym}

The fundamental field of Yang-Mills theory is a connection or vector potential
$A_\mu^{(a)}$ which in addition to the spacetime index $\mu$ carries an 
index $(a)$ referring to the internal gauge group.
We use internal indices from the beginning of the Latin alphabet 
$a,b,\ldots$ ranging over $1,2,\ldots,N$, where $N$ is the dimension of the
associated Lie algebra, e.g. $N=3$ if the gauge group is $SU(2)$.
Repeated internal indices are summed over.

Since the Yang-Mills equations are conformally invariant, we choose to work 
directly in the conformal spacetime here.
The Yang-Mills connection and its conformal counterpart are identical,
$A_\mu^{(a)} = \tA_\mu^{(a)}$.
We regard
\begin{equation}
  \tA_0^{(a)} \equiv - \tpsi^{(a)} 
\end{equation}
as a gauge variable that can be freely chosen.

The Yang-Mills field strength tensor is defined as
\begin{equation}
  \label{e:ymF}
  \tF_{\mu\nu}^{(a)} = \partial_\mu \tA_\nu^{(a)} - \partial_\nu \tA_\mu^{(a)}
    + f^{abc} \tA_\mu^{(b)} \tA_\nu^{(c)},
\end{equation}
where the symbol $f^{abc}$ is totally antisymmetric in the chosen basis of the 
Lie algebra.
For $SU(2)$ we may write $f^{abc} = g [abc]$, where $g$ is a constant;
we set $g=-2$ for our numerical evolutions in section \ref{s:numresults}.
The symbol $[abc]$ is totally antisymmetric with $[123] = 1$.

We introduce the electric field
\begin{equation}
  \tcD^{i(a)} \equiv \sqrt{-\four{\gamma}}\, \tF^{0i(a)} = \tN \mu_\gamma \tF^{0i(a)}
\end{equation}
and magnetic field
\begin{equation}
  \tcB^{i(a)} = \half \mu_\gamma \epsilon^{ijk} \tB_{jk}^{(a)} 
   = \half [ijk] \tF_{jk}^{(a)},
\end{equation}
where 
\begin{equation}
  \label{e:ymB}
  \tB_{ij}^{(a)} \equiv \tF_{ij}^{(a)}
\end{equation}
denotes the spatial part of the field strength tensor and 
$\epsilon_{ijk}$ is the alternating symbol associated with the conformal 
spatial metric $\gamma_{ij}$.
Note that $\tcD^{i}$ and $\tcB^i$ are vector densities of weight one.

The energy-momentum tensor is given by
\begin{equation}
  \tT_{\mu\nu} = \tF_{\mu\rho}^{(a)} \tF_\nu{}^{\rho(a)} 
     - \fourth \four{\gamma}_{\mu\nu} \tF_{\rho\sigma}^{(a)} \tF^{\rho\sigma(a)}.
\end{equation}
The projections defined in \eref{e:cTTprojs} take the form
\begin{eqnarray}
  \tilde \rho = \half \mu_\gamma^{-2} (\tcD^{i(a)} \tcD_i^{(a)} 
    + \tcB^{i(a)} \tcB_i^{(a)}),\\
  \tilde J_i = \mu_\gamma^{-1} [ijk] \tcD^{j(a)} \tcB^{k(a)},\\
  \tilde S = \tilde \rho,\\
  \tilde S^{\tr ij} = -\mu_\gamma^{-2} [\tcD^{i(a)} \tcD^{j(a)} 
    + \tcB^{i(a)} \tcB^{j(a)}]^{\tr}.
\end{eqnarray}

The Yang-Mills equations are
\begin{equation}
  \label{e:ym}
  \ftnabla_\mu \tF^{\mu\nu(a)} + f^{abc} \tA_\mu^{(b)} \tF^{\mu\nu(c)} = 0.
\end{equation}
Contracting this with $\tn_\nu$, we obtain the constraint
\begin{equation}
  \label{e:ymconstr}
  \partial_i \tcD^{i(a)} + f^{abc} A_i^{(b)} \tcD^{i(c)} = 0;
\end{equation}
contracting with $\gamma_\nu{}^i$ instead,
\begin{eqnarray}
  \label{e:ymdtD}
  \fl \partial_t \tcD^{i(a)} - (X^j \tcD^{i(a)})_{,j} + X^i{}_{,j} \tcD^{j(a)} =
    \nonumber\\ 
    \partial_j(\tN\mu_\gamma \tB^{ij(a)}) + \tN\mu_\gamma f^{abc} \tA_j^{(b)} 
    \tB^{ij(c)} + f^{abc} (\psi^{(b)} + X^k A_k^{(b)}) \tcD^{i(c)}.
\end{eqnarray}
An evolution equation for the vector potential is obtained from the definition
of $\tF_{0i}^{(a)}$,
\begin{equation}
  \label{e:ymdtA}
  \partial_t \tA_i^{(a)} = -\tN \mu_\gamma^{-1} \tcD_i^{(a)} - X^j \tB_{ij}^{(a)}
    - \tpsi^{(a)}_{,i} - f^{abc} \tA_i^{(b)} \tpsi^{(c)}.
\end{equation}
Equations \eref{e:ymdtA} and \eref{e:ymdtD} constitute the evolution equations
for the independent variables $\tA_i^{(a)}$ and $\tcD^{i(a)}$.
Note that $\tB_{ij}^{(a)}$ is to be expressed in
terms of $\tA_i^{(a)}$ and its spatial derivatives using 
definitions \eref{e:ymF} and \eref{e:ymB}.

%%%%%%%%%%%%%%%%%%%%%%%%%%%%%%%%%%%%%%%%%%%%%%%%%%%%%%%%%%%%%%%%%%%%%%%%%%%%%%%
%%%%%%%%%%%%%%%%%%%%%%%%%%%%%%%%%%%%%%%%%%%%%%%%%%%%%%%%%%%%%%%%%%%%%%%%%%%%%%%

\section{Reduction to spherical symmetry}
\label{s:spher}

In this section we reduce the formulation of the Einstein equations 
presented in section \ref{s:generalsources} and the specific matter models
of section \ref{s:matter} to spherical symmetry.
Special care is needed in the case of Yang-Mills theory.

\subsection{Einstein equations in isotropic gauge}

We work in spherical polar coordinates $t, r, \theta, \varphi$.
All fields are functions of $t$ and $r$ only. 
Partial derivatives w.r.t. time and radius are denoted by 
$\dot{} = \partial_t$ and $' = \partial_r$.

In spherical symmetry we may take the spatial conformal metric to be flat,
\begin{equation}
  \label{e:gammaflat}
  \gamma_{ij} = \mathrm{diag}(1, r^2, r^2 \sin^2\theta).
\end{equation}
The shift vector has a radial component only, and we set $X^r \equiv r X$.
The traceless momentum is diagonal and has only one independent component,
\begin{equation}
  \pi^{\tr rr} = -2r^2 \pi^{\tr \theta\theta} = -2r^2 \sin^2\theta\,
  \pi^{\tr\varphi\varphi} \equiv r^2 \sin\theta \, \hat \pi.
\end{equation}
(We have pulled out a factor of $\mu_\gamma = r^2 \sin\theta$ in the 
definition of the variable $\hat \pi$ in order to avoid frequent divisions by 
$\mu_\gamma$ in the final equations.)

Preservation of \eref{e:gammaflat} under the evolution equation \eref{e:dtgamma}
implies the spatial gauge condition
\begin{equation}
  \label{e:spher_isotropic}
  rX' = -\tfrac{3}{2} \tN \hat \pi.
\end{equation}
This implies that $\hat \pi$ is actually $\Or(r^2)$ near the origin 
and so we set 
\begin{equation}
  \hat \pi \equiv r^2 \pi.
\end{equation}
We also note the expression for the conformal mean curvature in our gauge,
\begin{equation}
  C = \tN^{-1} \tnabla_i X^i
  = -\tfrac{3}{2} r^2 \pi + 3 \tN^{-1} X.
\end{equation}

The reduction of the Einstein equations is as follows.
The Hamiltonian constraint \eref{e:hamcons} reads
\begin{eqnarray}
  \label{e:spher_hamcons}
 -4 \Omega \Omega'' + 6 \Omega'^2 
  - 8 \Omega r^{-1}\Omega' + \tfrac{3}{2} \Omega^2 r^4 \pi^2 - \tfrac{2}{3} K^2
  + 2 \kappa \Omega^4 \tilde \rho = 0.
\end{eqnarray}
The momentum constraint \eref{e:momcons} is
\begin{equation}
  \label{e:spher_momcons}
  \Omega (r\pi' + 5 \pi) - 2r\Omega' \pi 
  + \kappa \Omega^3 r^{-1} \tilde J^r = 0.
\end{equation}
The CMC slicing condition \eref{e:cmc} is
\begin{eqnarray}
  \label{e:spher_cmc}
  \fl -\Omega^2 \tN'' + 3 \Omega\Omega'\tN' 
  - 2\Omega^2 r^{-1}\tN' - \tfrac{3}{2}\Omega'^2 \tN + \sixth \tN K^2 
  + \tfrac{15}{8} \tN \Omega^2 r^4 \pi^2\nonumber\\
  + \half \kappa \tN \Omega^4 (\tilde S + 2 \tilde \rho) = 0.
\end{eqnarray}
The evolution equation for the conformal factor \eref{e:dtOmega} is
\begin{equation}
  \label{e:spher_dtOmega}
  \dot \Omega = r X \Omega' - X \Omega + \tN(\half \Omega r^2 \pi - \third K)
\end{equation}
and that for the traceless momentum \eref{e:dtpi} 
\begin{eqnarray}
  \label{e:spher_dtpi}
  \fl \dot \pi = r X \pi' + 3 X \pi + \tfrac{2}{3} r^{-1}(r^{-1}\tN')'\nonumber\\
  + \tN \left[ -\tfrac{4}{3} \Omega^{-1} r^{-1} (r^{-1}\Omega')'
    - \tfrac{2}{3} \Omega^{-1} K \pi - \half r^2 \pi^2
   + \kappa \Omega^2 r^{-2} \tilde S^{\tr rr}\right].
\end{eqnarray}

\subsubsection{Regularity at the origin.}

The fields $\Omega$, $\pi$, $\tN$ and $X$ are even functions of $r$.
It follows from the general behaviour near the origin of spherically symmetric
tensor fields \cite{Rinne2005,Ruiz2007a} that the $r$-component of a vector
$\tilde J^i$ is $\Or(r)$ and the $rr$-component of a tracefree symmetric tensor 
$\tilde S^{\tr ij}$ is $\Or(r^2)$ near the origin; this can also be seen 
explicitly for the specific matter models in the following subsections.
Hence the terms $\rho$, $r^{-1} \tilde J^r$, $\tilde S$ and 
$r^{-2} \tilde S^{\tr rr}$ appearing as sources in 
\eref{e:spher_hamcons}--\eref{e:spher_dtpi} are also regular even functions 
of $r$.
Keeping this in mind, equations \eref{e:spher_hamcons}--\eref{e:spher_dtpi}
have been written in a form that is manifestly regular at the origin.

\subsubsection{Regularity at future null infinity.}
\label{s:spher_scrireg}

For completeness and for later use we also state the
results of our regularity analysis at \scri here.
In the following $\hateq$ denotes equality at \scri.
By definition,
\begin{equation}
  \label{e:Omega_scri}
  \Omega \hateq 0.
\end{equation}
The Hamiltonian constraint \eref{e:spher_hamcons} and momentum constraint 
\eref{e:spher_momcons} yield
\begin{eqnarray}
  \Omega' \hateq r \Omega'' \hateq -\third K, \quad \Omega''' \hateq 0,\\
  \label{e:pireg}
  \pi \hateq \pi' \hateq 0.
\end{eqnarray}
From the CMC condition  \eref{e:spher_cmc} we obtain
\begin{equation}
  r \tN' \hateq r^2 \tN'' \hateq \tN.
\end{equation}
Note that the value of $\tN$ at \scri can be freely chosen; see section
\ref{s:bcs} for our particular choice.
Preservation of \eref{e:Omega_scri} under the evolution equation
\eref{e:spher_dtOmega} implies
\begin{equation}
  \label{e:X_scri}
  r X \hateq -\tN.
\end{equation}
The isotropic gauge condition \eref{e:spher_isotropic} further yields
\begin{equation}
  X' \hateq X'' \hateq 0.
\end{equation}
Given these results it can be verified using L'Hospital's rule that the 
evolution equation \eref{e:spher_dtpi} for the traceless momentum
reduces to
\begin{equation}
  \partial_t \pi \hateq 0,
\end{equation}
consistent with \eref{e:pireg}.

%%%%%%%%%%%%%%%%%%%%%%%%%%%%%%%%%%%%%%%%%%%%%%%%%%%%%%%%%%%%%%%%%%%%%%%%%%%%%%%%

\subsection{The conformally coupled scalar field}

The scalar field evolution equations \eref{e:cc_dtphi}--\eref{e:cc_dtpsi}
take the form
\begin{eqnarray}
  \label{e:spher_dtphi}
  \fl \dot {\tphi} = r X \tphi' + X \tphi + \tN(-\half \tphi r^2 \pi + \tpsi),\\
  \label{e:spher_dtpsi}
  \fl \dot {\tpsi} = (r X \tpsi + \tN \tphi')' + X\tpsi + \tN( 2r^{-1}\tphi'
  - \fourth \tphi r^4 \pi^2 + \half \tpsi r^2 \pi)
  + \third \tphi (\tN'' + 2r^{-1}\tN').
\end{eqnarray}

The source terms \eref{e:cc_rho}--\eref{e:cc_Str} appearing in the Einstein 
equations are 
\begin{eqnarray}
  \fl \tilde \rho = -\half \tphi(\tphi'' + 2r^{-1}\tphi') + \fourth \tphi'^2 
     + \tfrac{3}{4}\tpsi^2 - \tfrac{3}{16}\tphi^2 r^4 \pi^2,\\
  \label{e:scalarJr}   
  \fl r^{-1}\tilde J^r = -r^{-1}\tphi'\tpsi + \half \tphi r^{-1}\tpsi' 
     - \half \tphi\tphi' r\pi - \fourth\tphi^2(r\pi' + 5\pi),\\
  \fl \tilde S = \tilde \rho,\\
  \fl r^{-2}\tilde S^{\tr rr} = -\third \tphi r^{-1}(r^{-1}\tphi')' 
     + \tfrac{2}{3} (r^{-1}\tphi')^2  + \half \tphi\tpsi\pi \nonumber\\
     + \fourth \tphi^2\left[ \tN^{-1}(\dot \pi - r X \pi' - 3X\pi)
        + \half r^2 \pi^2 - \tfrac{2}{3}\tN^{-1} r^{-1}(r^{-1}\tN')'\right].
\end{eqnarray}

Notice again that these equations are manifestly regular at the origin, given 
that $\tphi$ and $\tpsi$ are even functions of $r$.

%%%%%%%%%%%%%%%%%%%%%%%%%%%%%%%%%%%%%%%%%%%%%%%%%%%%%%%%%%%%%%%%%%%%%%%%%%%%%%%%

\subsection{Yang-Mills theory}

In the case of Yang-Mills it is more convenient to work in Cartesian 
coordinates, so indices $i,j,\ldots$ will refer to Cartesian coordinates in
this subsection.
Hence the conformal metric is now
$\gamma_{ij} = \delta_{ij}$, with $\mu_\gamma = 1$, and the shift vector is 
$X^i = X x^i$.

We take the gauge group to be $SU(2)$.
The most general ansatz for the spherically symmetric Yang-Mills connection 
is of the form \cite{Witten1977,Gu1981,SarbachPhD}
\begin{equation}
  \label{e:spherym}
  \tA^{i(a)} = [aij] x^j F + (x^a x^i - r^2 \delta^{ai}) H + \delta^{ai} L ,
  \qquad \tA_0^{(a)} = G x^a.
\end{equation}
Here $F, H, L$ and $G$ are functions of $t$ and $r$ only.

Similarly, we write the electric field as
\begin{equation}
  \tcD^{i(a)} = [aij] x^j D_F + (x^a x^i - r^2 \delta^{ai}) D_H + \delta^{ai} D_L.
\end{equation}
The evolution equations for the vector potential \eref{e:ymdtA} read
\begin{eqnarray}
  \label{e:dtF}
  \dot F &=& rXF' - \tN D_F + 2XF + g(G-XL)(L-r^2 H),\\
  \label{e:dtH}
  \dot H &=& rXH' - \tN D_H + r^{-1} G' + g (G-XL)F + X(3H - r^{-1} L'),\\
  \label{e:dtL}
  \dot L &=& -\tN D_L + rG' + G.
\end{eqnarray}
The evolution equations for the electric field \eref{e:ymdtD} are
\begin{eqnarray}
   \label{e:dtDF}
  \fl \dot D_F =(rX D_F - \tN F')' + 2X D_F + g(D_L - r^2 D_H)(G-XL)\nonumber\\
    + \tN [-4r^{-1} F'- 2gLrH' + gr^{-1} L'(3L - r^2H) ]\nonumber\\
    + r^{-1} \tN' [-2F + gL(L-r^2 H)]
    + \tN g[ - 3F^2 - r^2H^2 - 4HL\nonumber\\
    \qquad+ g(r^2F^3 - 2r^2FHL + r^4FH^2 + 2FL^2)],\\
    \label{e:dtDH} 
  \fl \dot D_H = (rXD_H - \tN H')' - r^{-1}(XD_L - \tN r^{-1} L')'
    + r^{-1} \tN'(-3H + gFL) \nonumber\\
    + gD_F(G + XL - 2Xr^2H) + XD_H(1 + 2gr^2F) - 2gXFD_L\nonumber\\
    +\tN[-4r^{-1} H' + g(2HrF' - 2FrH' + 3Fr^{-1} L')]\nonumber\\
    + \tN g\{-4FH + g[F^2(L + r^2H) + r^4 H^3 - r^2H^2L]\},\\
    \label{e:dtDL}
  \fl \dot D_L = X[2(1-gr^2F)(D_L - r^2 D_H) + 2gr^2 D_F(L-r^2 H)]\nonumber\\
    + 2 \tN (1-gr^2F)(3H + rH' - r^{-1} L' - gFL)\nonumber\\
    - 2\tN g(L-r^2H)[rF' - gr^2 H(L-r^2 H) + 2F - g(L-r^2H)^2]. 
\end{eqnarray}
The Yang-Mills constraint \eref{e:ymconstr} is
\begin{equation}
  \label{e:spher_YMconstr}
  r^{-1} D_L' + 2gF(D_L - r^2 D_H) + 2 D_H + 2g(r^2H - L) D_F = 0 .
\end{equation}
The components of the magnetic field
\begin{equation}
  \tcB^{i(a)} = [aij] x^j B_F + (x^a x^i - r^2 \delta^{ai}) B_H + \delta^{ai} B_L
\end{equation} 
evaluate to
\begin{eqnarray}
  \label{e:BF}
  B_F &=& -3H - rH' + r^{-1} L' + gFL,\\
  \label{e:BH}
  B_H &=& r^{-1} F' - gH(L-r^2 H) + gF^2,\\
  \label{e:BK}
  B_L &=& -2F + g(L-r^2 H)^2 + gr^2F^2.
\end{eqnarray}
In terms of these, the matter sources in the Einstein equations are given by
\begin{eqnarray}
  \fl \tilde \rho = \tilde S = \half[3D_L^2 - 2r^2(2D_L D_H - D_F^2 - r^2 D_H^2)
     \nonumber\\+ 3 B_L^2 - 2r^2(2 B_L B_H - B_F^2 - r^2 B_H^2)],\\
  \fl r^{-1} \tilde J^r = 2[D_L B_F - D_F B_L + r^2(D_F B_H - D_H B_F)],\\
  \fl r^{-2}\tilde S^{\tr rr} = -\tfrac{2}{3} (2D_L D_H - D_F^2 - r^2 D_H^2 
     + 2 B_L B_H - B_F^2 - r^2 B_H^2). 
\end{eqnarray}

We are free to impose one gauge condition on the Yang-Mills connection, and we
choose \emph{radial gauge} $L = 0$.
The evolution equation \eref{e:dtL} for $L$ now turns into an ODE determining
$G$,
\begin{equation}
  \label{e:YMradial}
  rG' + G - \tN D_L = 0.
\end{equation}

If the initial data are such that $H = D_H = 0$ then $D_L = 0$ is a solution 
to \eref{e:spher_YMconstr}, and $G=0$ (i.e.~temporal gauge) is a solution to 
\eref{e:YMradial}.
(There are other solutions to these last two equations but the boundary
conditions we impose will single out the given ones.)
These conditions are preserved under the time evolution, 
i.e.~$H = G = D_H = D_L = 0$ at all times.
This reduced system is sometimes referred to as \emph{purely magnetic} or as
the \emph{gravitational sector} of the Einstein-Yang-Mills equations, 
whereas the remaining system is called the 
\emph{sphaleron sector} \cite{Choptuik1999}.
While the gravitational sector has been widely studied 
numerically (e.g.~in \cite{Choptuik1996,Choptuik1999,Puerrer2009}),
we are not aware of any numerical simulations using the full system.

We also note that the ansatz \eref{e:spherym} is often written (e.g. in 
\cite{Choptuik1999}) in a different gauge known as \emph{Abelian gauge},
\begin{equation}
  \label{e:abeliangauge}
  \tilde A = w \tau^\theta \rmd\theta + (\cot \theta \, \tau^r + w \tau^\phi)
  \sin \theta \, \rmd \theta,
\end{equation}
where $\tau^i$ denote the generators of $SU(2)$.
The two ans\"atze are related by a (singular) $SU(2)$ gauge transformation 
(see e.g. \cite{Fodor2008}).

%%%%%%%%%%%%%%%%%%%%%%%%%%%%%%%%%%%%%%%%%%%%%%%%%%%%%%%%%%%%%%%%%%%%%%%%%%%%%%%
%%%%%%%%%%%%%%%%%%%%%%%%%%%%%%%%%%%%%%%%%%%%%%%%%%%%%%%%%%%%%%%%%%%%%%%%%%%%%%%

\section{Numerical method}
\label{s:nummethod}

In this section we describe the numerical methods we use in order to solve the
spherically symmetric reduction of the Einstein equations on hyperboloidal
slices derived in section \ref{s:spher}. 
We begin by summarising our evolution scheme.
Particular care is spent on the boundary conditions.
We then describe the finite-difference discretisation, time integration method
and elliptic solver.
Finally we explain how apparent horizons are detected and excised.
The code has been implemented in Python.

\subsection{Evolution scheme}
\label{s:evolscheme}

Our fundamental geometric variables are $\Omega$, $\pi$, $\tN$ and $X$.
The matter variables are either $\tphi$ and $\tpsi$ for the scalar field,
or $F$, $H$, $G$, $D_F$, $D_H$ and $D_L$ for Yang-Mills.

The numerical evolution proceeds as follows.
At each time step, we solve
\begin{enumerate}
  \label{i:ymconstr}
  \item the Yang-Mills constraint \eref{e:spher_YMconstr} for $D_L$,
  \item the Einstein constraints \eref{e:spher_hamcons} and 
    \eref{e:spher_momcons} for $\Omega$ and $\pi$.
    If the source $\tilde J^r$ of the momentum constraint is independent of
    $\pi$, as is the case for Yang-Mills, then the two equations may be
    decoupled by setting $\pi = \Omega^2 P$, which turns the momentum 
    constraint \eref{e:spher_momcons} into
    \begin{equation}
      r P' + 5 P + \kappa r^{-1} \tilde J^r = 0.
    \end{equation}
    This is solved first and the solution for $P$ is then substituted in the
    Hamiltonian constraint \eref{e:spher_hamcons}.
    For the conformally coupled scalar field, the two constraints cannot be 
    decoupled in this way because $\tilde J^r$ depends on $\pi$
    (cf.~equation \eref{e:scalarJr}).
    In this case the Einstein constraints must be solved as a coupled system.
  \item the slicing condition \eref{e:spher_cmc} for $\tN$,
  \item the isotropic spatial gauge condition \eref{e:spher_isotropic} for $X$,
  \item the Yang-Mills radial gauge condition \eref{e:YMradial} for $G$.
\end{enumerate}
(Obviously steps (i) and (v) are only included for Yang-Mills matter.)
The matter variables are then evolved to the next time step using either
\eref{e:spher_dtphi} and \eref{e:spher_dtpsi} for $\tphi$ and $\tpsi$ or
\eref{e:dtF}, \eref{e:dtH}, \eref{e:dtDF} and \eref{e:dtDH} for $F$, $H$, 
$D_F$ and $D_H$.

The evolution equations \eref{e:spher_dtOmega}, \eref{e:spher_dtpi} and
\eref{e:dtDL} for $\Omega$, $\pi$ and $D_L$ are not solved explicitly but
are evaluated during the numerical evolution as a consistency check.

%%%%%%%%%%%%%%%%%%%%%%%%%%%%%%%%%%%%%%%%%%%%%%%%%%%%%%%%%%%%%%%%%%%%%%%%%%%%%%%%

\subsection{Boundary conditions}
\label{s:bcs}

We discuss the boundary conditions for each of the equations in turn.
We need to distinguish between two cases, a regular centre and
a black hole with an inner excision boundary.
The outer boundary is always placed at \scri.

The Yang-Mills constraint \eref{e:spher_YMconstr} is a first-order ODE 
for $D_L$ that requires one Dirichlet boundary condition. 
The value of $D_L$ at either the origin or the excision boundary is obtained
by evolving $D_L$ there according to its evolution equation \eref{e:dtDL}.

The Hamiltonian constraint \eref{e:spher_hamcons} is a second-order ODE for 
$\Omega$, and we regard it as a two-point boundary value problem.
For a regular centre, the inner boundary condition follows from the 
fact that $\Omega$ is an even function of $r$: $\Omega' = 0$ at $r=0$.
For an excised black hole, we use the evolution equation \eref{e:spher_dtOmega}
to find the value of $\Omega$ at the excision boundary.
The outer boundary condition at \scri is $\Omega \hateq 0$ in both cases.

The momentum constraint \eref{e:spher_momcons} is a first-order ODE for $\pi$.
At an excision boundary we use the evolution equation \eref{e:spher_dtpi}
to obtain a Dirichlet boundary condition for $\pi$ there.
For a regular centre, however, there is a unique solution 
to \eref{e:spher_momcons} that is regular at $r=0$.
This can be seen immediately by evaluating \eref{e:spher_momcons} at $r=0$, 
where the equation implies a Dirichlet condition for $\pi$.
Thus no additional boundary condition is imposed at $r=0$.

The CMC slicing condition \eref{e:spher_cmc} is a second-order ODE for $\tN$
that we regard as a two-point boundary value problem.
At a regular origin $r=0$ the boundary condition is $\tN' = 0$.
At an excision boundary we choose to freeze the value of $\tN$ from the
time when the black hole is excised.
Consider the proper time $\tau$ of an observer who remains at a fixed spatial 
coordinate location.
This is given by
\begin{equation}
  -\rmd \tau^2 = \Omega^{-2}(-\tN^2 + r^2 X^2)\rmd t^2.
\end{equation}
Using the results of our regularity analysis (section \ref{s:spher_scrireg}), 
the formally singular term on the right-hand side can be shown to have a 
regular limit at \scri,
\begin{equation}
  \Omega^{-2} (\tN^2 - r^2 X^2) \hateq 9 r^{-2} K^{-2} \tN^2.
\end{equation}
Hence choosing $t$ to coincide with $\tau$ at \scri corresponds to setting
\begin{equation}
  \tN \hateq \third K r.
\end{equation}

The isotropic spatial gauge condition \eref{e:spher_isotropic} is a first-order
ODE for $X$. 
The boundary condition for $X$ at \scri follows from preservation of 
$\Omega \hateq 0$ under the evolution equation \eref{e:spher_dtOmega},
namely, $X \hateq - r^{-1} \tN$.

Finally we consider the Yang-Mills radial gauge condition \eref{e:YMradial}.
This first-order ODE for $G$ has a unique solution that is regular at $r=0$. 
(Evaluating the equation at $r=0$ implies a Dirichlet condition for $G$.)
Thus no additional boundary condition must be imposed at $r=0$.
For an inner excision boundary we choose to freeze the value of $G$ there
from the time of excision.

Since the outer boundary at \scri is null and the inner excision boundary is
spacelike, the evolution equations (in particular those for the matter 
variables) do not require any boundary conditions there.
Near a regular origin the evolution equations need not be modified either.
The way we discretise the equations near the boundaries is described in the
following subsection.

%%%%%%%%%%%%%%%%%%%%%%%%%%%%%%%%%%%%%%%%%%%%%%%%%%%%%%%%%%%%%%%%%%%%%%%%%%%%%%%%

\subsection{Discretisation}

The numerical domain is an interval $r \in [0, 1]$ for a regular centre
or $[\rmin, 1]$ for an excised centre. 
In both cases the outer boundary at $r=1$ corresponds to \scri.

For solutions containing a black hole, the fields typically have
steep gradients close to the inner (excision) boundary and hence it is 
advisable to introduce a non-uniform grid.
We do this by introducing a new radial coordinate $x$
with respect to which the numerical grid is uniform, combined with a map
\begin{equation}
  r: [0,1] \rightarrow [\rmin, 1], \quad x \mapsto r(x),
\end{equation}
that is steeper near $x=1$ than near $x=0$, thus providing more resolution
close to the inner boundary.
As in \cite{Rinne2010} we choose
\begin{equation}
  \label{e:excmap}
  r(x) = Q_1 x^2 + (1-\rmin - Q_1)x + \rmin,
\end{equation}
where $0\leqslant Q_1 < 1$ is a constant, typically $Q_1=0.5$.
For evolutions with a regular centre, we use $r(x) = x$ for the numerical 
results shown here.
More generally, we have implemented a map
\begin{equation}
  \label{e:regmap}
  r(x) = Q_2 x + (1-Q_2)x^3,
\end{equation}
with $0\leqslant Q_2 \leqslant 1$ a constant.
Note that since $r(x)$ in \eref{e:regmap} is an odd function of $x$, 
the parity of a grid function with respect to $x$ will be the same as with 
respect to $r$.
(This is not the case for \eref{e:excmap} but parity considerations are
only relevant for regularity at the origin.)

The interval $x\in [0,1]$ is covered by equidistant grid points.
For a regular centre we use a grid that is staggered about the origin,
\begin{equation}
  \label{e:staggeredgrid}
  x_i = (i+\half) h, \quad 0\leqslant i \leqslant N,
\end{equation}
whereas for an excised centre we use an unstaggered grid
\begin{equation}
  x_i = i h, \quad 0\leqslant i \leqslant N,
\end{equation}
so that $r_0 \equiv r(x_0) = r(0) = \rmin$ lies on the excision boundary.
In both cases, the grid is chosen such that $r_N \equiv r(x_N) = 1$.

Derivatives with respect to $x$ are discretised using fourth-order accurate
finite differences. 
Explicit expressions for the finite-difference operators can be found in
Appendix C of \cite{Rinne2010}.
We use one-sided differences near \scri and the inner excision boundary.
For a regular centre, the usual centred finite-difference operators are used 
near the origin by formally extending the staggered grid \eref{e:staggeredgrid}
to negative values of $i$ and using the fact that all the evolved variables are
even functions of $r$ (and hence of $x$), i.e. replacing
\begin{equation}
  \label{e:ghosts}
  x_{-1} = x_0, \quad x_{-2} = x_1, \quad x_{-3} = x_2, \quad \ldots \quad .
\end{equation}

The reader will have noticed that we have written the highest derivatives 
in the matter evolution equations \eref{e:spher_dtpsi} and
\eref{e:dtDF}--\eref{e:dtDH} in flux-conservative form. 
When discretising these equations, we apply the finite-difference operators in
prescisely the order in which the terms in the continuum equations are written.
This has been found to be essential for the stability of the method.

%%%%%%%%%%%%%%%%%%%%%%%%%%%%%%%%%%%%%%%%%%%%%%%%%%%%%%%%%%%%%%%%%%%%%%%%%%%%%%%%

\subsection{Time integration}

A fourth-order Runge-Kutta method is used in order to integrate the evolution
equations forward in time.
At each full timestep, the elliptic equations are solved as described
below in section \ref{s:elliptic}.
At the substeps of the Runge-Kutta algorithm, we have found it sufficient
to extrapolate the solution to the elliptic equations from previous timesteps
(using a cubic polynomial approximation) instead of solving the equations
explicitly.
Kreiss-Oliger dissipation \cite{Kreiss1973} is added in order to ensure 
stability; see Appendix C of \cite{Rinne2010}.
For a regular centre, the dissipation operator is evaluated up to the
innermost grid point using the rule \eref{e:ghosts}.
No dissipation is added at the outermost two grid points near \scri and 
the excision boundary.

%%%%%%%%%%%%%%%%%%%%%%%%%%%%%%%%%%%%%%%%%%%%%%%%%%%%%%%%%%%%%%%%%%%%%%%%%%%%%%%%

\subsection{Elliptic solver}
\label{s:elliptic}

The elliptic equations listed in section \ref{s:evolscheme}, which are
in fact ODEs in spherical symmetry, are solved using a combination of a
Newton-Raphson iteration and a direct band-diagonal solver.
The Newton-Raphson iteration is actually only needed for the Hamiltonian
constraint \eref{e:spher_hamcons}, all the other equations are linear.
The matrices arising from our fourth-order finite-difference discretisation
are pentadiagonal, except near \scri and the excision boundary, where
one-sided differences with wider stencils are used; 
a few Gaussian eliminations are applied by hand in order to reduce the matrix
to pentadiagonal form there.
We use the Python routine {\tt numpy.linalg.solve\_banded} (a straightforward 
generalisation of the Thomas algorithm)
in order to solve this pentadiagonal linear system.

%%%%%%%%%%%%%%%%%%%%%%%%%%%%%%%%%%%%%%%%%%%%%%%%%%%%%%%%%%%%%%%%%%%%%%%%%%%%%%%%

\subsection{Apparent horizon finder, black hole excision and 
  Bondi mass}

We detect the formation of an apparent horizon by tracking the optical scalars
$\theta_\pm$ during the evolution.
These are defined as follows.
Let $n_\mu = \Omega^{-1} \tN (\rmd t)_\mu$ denote the (physical) unit timelike 
normal to a given $t=\const$ slice and 
$s^\mu = \Omega (\partial/\partial r)^\mu$ 
the unit outward normal to an $r=\const$ surface within that slice.
The outgoing and ingoing null normals to this surface are 
$l_\pm^\mu = n^\mu \pm s^\mu$.
The optical scalars or null expansions are now given by
\begin{equation}
  \label{e:opticalscalars}
  \theta_\pm = (\ln R)_{,\mu} l_\pm^\mu = \third K - \half \Omega r^2 \pi
     \pm (r^{-1} \Omega - \Omega'),
\end{equation}
where $R \equiv r/\Omega$ denotes (physical) areal radius.

An apparent horizon is given by the outermost radius $r$ at which 
$\theta_+ = 0$. 
A zero of this function is detected using a standard root finding algorithm
(the Python routine {\tt scipy.optimize.brentq}).
Even before a zero forms, we can detect and track a minimum of $\theta_+$
in order to have a better initial guess at where the zero is about to occur.

Once an apparent horizon has formed at a radius $r=\rAH$
(with corresponding mass $\MAH = \half R_\mathrm{AH}$),
we excise just inside it, at $r=0.9 \, \rAH \equiv \rmin$.
The numerical solution is interpolated (using cubic interpolation) 
to a new grid with inner boundary at $r=\rmin$, 
discarding the part of the solution at smaller values of $r$. 
The evolution is then continued on the new grid.

The optical scalars \eref{e:opticalscalars} are closely related to the
Hawking mass
\begin{equation}
  \label{e:Hawking}
  M_\mathrm{H} = \half R (1 + R^2 \theta_+ \theta_-).
\end{equation}
Its limit at \scri is the Bondi mass $M_\mathrm{B}$, which will be a useful
quantity to evaluate during our numerical evolutions.
Although formally singular, the results of our regularity analysis
(section \ref{s:spher_scrireg}) imply that the limit at \scri 
of \eref{e:Hawking} is
\begin{equation}
  \MB = M_\mathrm{H} |_{\mathrsfs{I}^+} 
   = -\tfrac{3}{4} K^{-2} (K r^5 \pi'' + 2 r^3 \Omega^{(4)}).
\end{equation}
Evaluating the fourth derivative of the conformal factor at \scri is prone to
numerical error;
we have found it to be more accurate to extrapolate $M_\mathrm{H}$ 
from the interior.

%%%%%%%%%%%%%%%%%%%%%%%%%%%%%%%%%%%%%%%%%%%%%%%%%%%%%%%%%%%%%%%%%%%%%%%%%%%%%%%
%%%%%%%%%%%%%%%%%%%%%%%%%%%%%%%%%%%%%%%%%%%%%%%%%%%%%%%%%%%%%%%%%%%%%%%%%%%%%%%

\section{Numerical results}
\label{s:numresults}

\subsection{Initial data}
\label{s:ini}

We start the evolution from initial data that are close to either Minkowski
or Schwarzschild spacetime.
The geometry variables ($\Omega$, $\pi$, $\tN$ and $X$) are first set according
to the respective vacuum solution, then initial data for the matter fields
are specified, and finally the constraints and elliptic gauge conditions
are re-solved for the geometry variables.

Schwarzschild spacetime in CMC coordinates is given 
by \cite{Brill1980,Malec2003}
\begin{equation}
   ds^2 = -\left(1-\frac{2M}{\bar r}\right) \rmd t^2 + \frac{1}{f^2} \,
   \rmd {\bar r}^2 - \frac{2a}{f} \, \rmd t \, \rmd \bar r 
   + {\bar r}^2 (\rmd\theta^2 + \sin^2\theta \, \rmd\phi^2),
\end{equation}
where
\begin{equation}
   f(\bar r) = \left(1-\frac{2M}{\bar r} + a^2\right)^{1/2}, \quad
   a(\bar r) = \frac{K \bar r}{3} - \frac{C}{{\bar r}^2}\, ,
\end{equation}  
and $M$ (mass), $K$ (mean curvature) and $C$ are constants.
The radial coordinate $\bar r$ needs to be transformed to a new radial 
coordinate $r$ such that the spatial metric is manifestly conformal to the 
flat metric in the new coordinates.
This yields the ODE
\begin{equation}
  \label{e:rbarode}
  \frac{\rmd r}{\rmd \bar r} = \frac{r}{\bar r f(\bar r)}.
\end{equation}
Since the physical radial coordinate $\bar r$ has infinite range, it is
more convenient to work with $s \equiv 1/\bar r$ and set
\begin{equation}
  A(s) \equiv \third K - C s^3, \quad
  F(s) \equiv s^2 - 2Ms^3 + A(s)^2.
\end{equation}  
Equation \eref{e:rbarode} now takes the form
\begin{equation}
  \label{e:sode}
  \frac{\rmd s}{\rmd r} = \frac{-F(s)^{1/2}}{r}.
\end{equation}
First we determine the $r$-coordinate of the horizon by numerical integration,
\begin{equation}
  r_\mathrm{H} = \exp \left( -\int_0^{s_H} F(s)^{-1/2} \rmd s \right),
\end{equation}
where $s_H = 1/(2M)$.
(Recall that we choose \scri, i.e. $\bar r=\infty \Leftrightarrow s=0$, 
to correspond to $r=1$.)
We place the excision boundary at $\rmin = 0.9 \, r_\mathrm{H}$.
The ODE \eref{e:sode} is now solved numerically on the interval
$r \in [\rmin, 1]$ with initial condition $s(1) = 0$.
(We use the Python routine {\tt scipy.integrate.odeint}.)

In terms of the numerically determined function $s(r)$, the geometry variables
are obtained as
\begin{equation}
  \Omega = r s, \quad
  \pi = 2 C r^{-3} s^2, \quad
  \tN = r F(s)^{1/2}, \quad
  X = C s^3 - \third K.
\end{equation}
For $M = C = 0$ this reduces to Minkowski spacetime (with $r \in [0,1]$),
\begin{equation}
  \Omega = \sixth K (1 - r^2), \quad
  \pi = 0, \quad 
  \tN = \sixth K (1 + r^2), \quad
  X = -\third K.
\end{equation}

For the matter fields we consider three types of initial data,
\begin{enumerate}
  \item \emph{scalar field}: specify initial data for $(\tphi, \tpsi)$,
    take the Yang-Mills field to vanish,
  \item \emph{gravitational-sector Yang-Mills}: 
    specify initial data for $(F, D_F)$, all other fields vanish,
  \item \emph{sphaleron-sector Yang-Mills}: specify initial data for $(H, D_H)$,
    take $F = D_F = \tphi = \tpsi = 0$, solve for $D_L$ and $G$.
\end{enumerate}
Note that in case (iii), $F=D_F=0$ initially but not during the evolution.
In each case, the initial data are chosen to be a Gaussian that is approximately
ingoing initially, e.g. for the pair $(\tphi, \tpsi)$,
\begin{equation}
  \tphi = A \exp \left( - \frac{(r-r_0)^2}{2\sigma^2} \right), \quad
  \tpsi = \tphi' + r^{-1} \tphi.
\end{equation}
For all the numerical evolutions shown here, we choose $r_0 = 0.5$ and 
$\sigma = 0.05$.

If we add these initial data for the matter fields to the Minkowski 
background solution (re-solving the constraints), we obtain evolutions 
that either \emph{disperse} or \emph{collapse} to a black hole, 
depending on the amplitude $A$.
If the background geometry is Schwarzschild, the matter will partly
\emph{accrete} onto the black hole.

%%%%%%%%%%%%%%%%%%%%%%%%%%%%%%%%%%%%%%%%%%%%%%%%%%%%%%%%%%%%%%%%%%%%%%%%%%%%%%%%

\subsection{Scalar field evolutions}

\begin{figure}
\centerline{\includegraphics[scale=0.24]{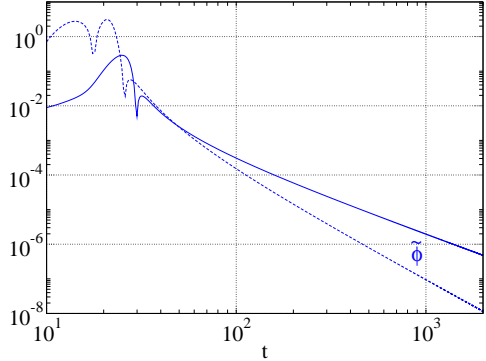}}
\caption{\label{f:scaldisp} 
  Scalar field dispersal. 
  The initial Bondi mass is $\MBi = 0.59$.
  At \scri (solid line) the field decays as $\tphi \sim t^{-2}$ at late times.
  At the origin (dashed line), $\tphi \sim t^{-3}$.
}
\end{figure}

We begin with a scalar field evolution that disperses to flat space
(figure \ref{f:scaldisp}).
The amplitude is chosen to be $A = 0.6$, which corresponds to an initial Bondi
mass $\MBi = 0.59$, already well in the non-linear regime.
At late times, the field decays as a power law.
At \scri, $\tphi \sim t^{-2}$, whereas at the origin and in fact at any finite 
radius, $\tphi \sim t^{-3}$.
These results are in agreement with \cite{Puerrer2005}.
The fact that the scalar field decays more slowly at \scri than away from
\scri leads to the solution becoming increasingly peaked at \scri, similar 
to the formation of a boundary layer.
Eventually this feature cannot be resolved at a fixed numerical resolution.
The simulations presented here were run at two different resolutions in 
order to make sure that the plots can be trusted during the times shown.
For the simulations shown in the plots, $N = 8000$ grid points were used.

\begin{figure}
\centerline{\includegraphics[scale=0.24]{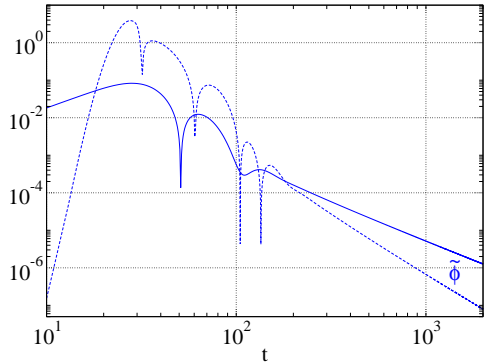}}
\caption{\label{f:scalaccr}
  Scalar field accretion. 
  The initial Bondi and apparent horizon masses are $\MBi = 1.45$, 
  $\MAHi = 1$, and the final masses are $\MBf = \MAHf = 1.44$.
  At \scri (solid line) the field decays as $\tphi \sim t^{-2}$ at late times.
  At the horizon (dashed line), $\tphi \sim t^{-3}$.
}
\end{figure}

Next we choose initial data containing a black hole with a scalar field 
perturbation (figure \ref{f:scalaccr}).
Initially the Bondi mass is $\MBi = 1.45$ and the apparent horizon mass is 
$\MAHi = 1$.
Almost all of the matter falls into the black hole---the final
Bondi mass, which agrees with the final apparent horizon mass, is 
$\MBf = \MAHf = 1.44$.
At late times we observe a power-law decay of the scalar field with the
same decay exponents as in the evolution that dispersed to flat space,
where now instead of the origin, we evaluate the field at the horizon.
These exponents agree with those found in the test field 
approximation \cite{Zenginoglu2008b}.

\begin{figure}
\centerline{\includegraphics[scale=0.24]{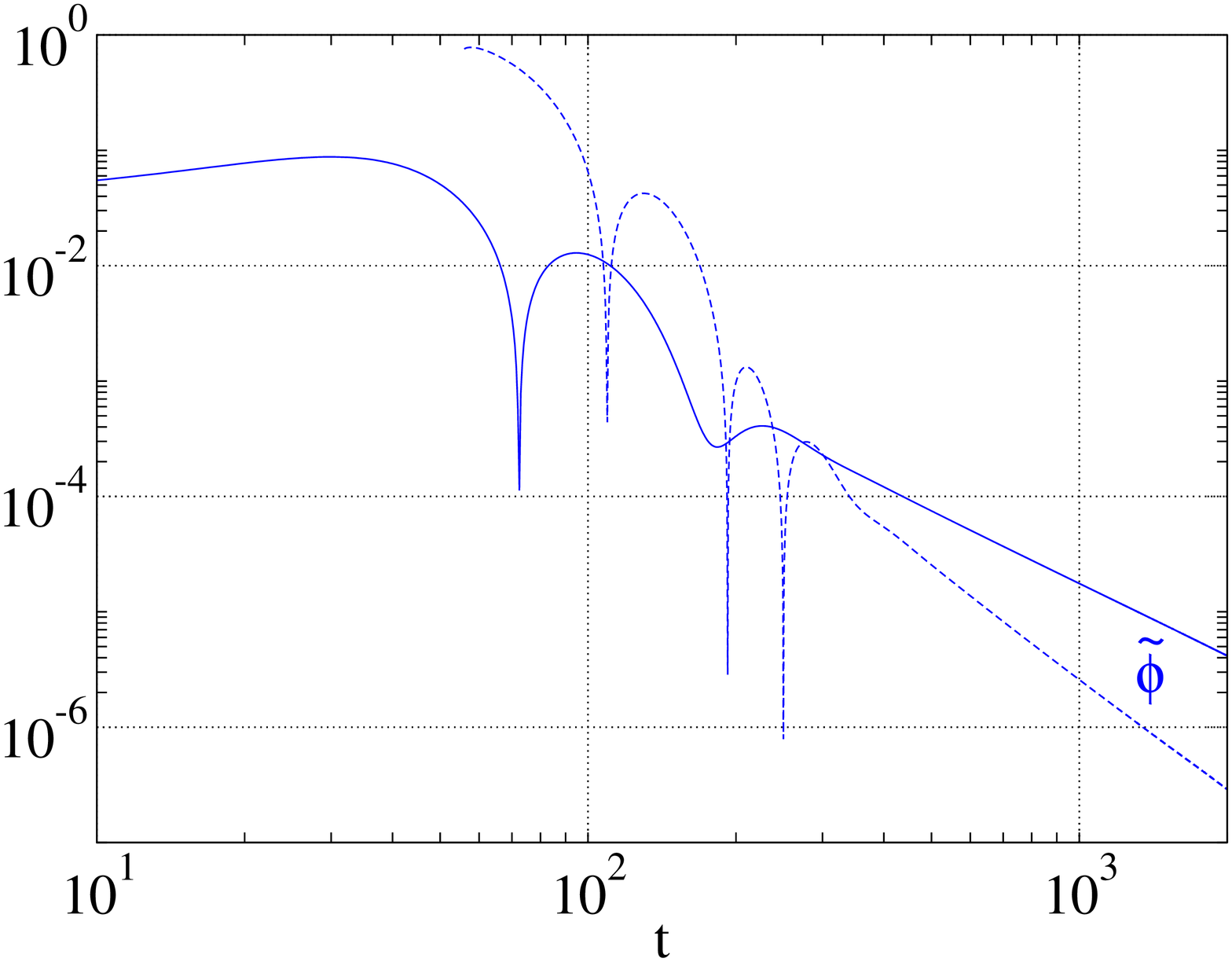}}
\caption{\label{f:scalcoll}
  Scalar field collapse. 
  The initial Bondi mass is $\MBi = 2.74$ and the final Bondi and 
  apparent horizon masses are $\MBf = \MAHf = 2.71$.
  At \scri (solid line) the field decays as $\tphi \sim t^{-2}$ at late times.
  At the horizon (after it forms, dashed line), $\tphi \sim t^{-3}$.
}
\end{figure}

Finally we return to the case of regular initial data as in the first simulation
but increase the amplitude such that a black hole forms in the course of the
evolution (figure \ref{f:scalcoll}).
The initial Bondi mass is $\MBi = 2.74$, and the final Bondi and apparent 
horizon masses are $\MBf = \MAHf = 2.71$.
Again we find the same decay exponents as in the dispersing evolution.

%%%%%%%%%%%%%%%%%%%%%%%%%%%%%%%%%%%%%%%%%%%%%%%%%%%%%%%%%%%%%%%%%%%%%%%%%%%%%%%%

\subsection{Gravitational-sector Yang-Mills evolutions}

\begin{figure}
\centerline{\includegraphics[scale=0.24]{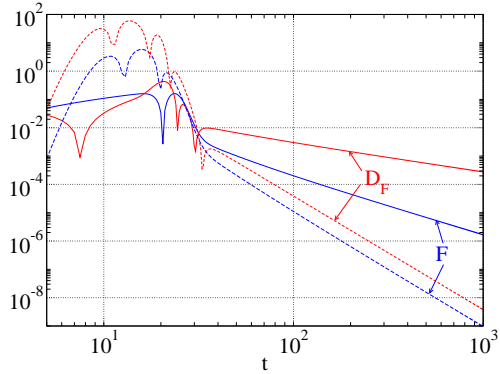}}
\caption{\label{f:ym1disp} 
  Gravitational-sector Yang-Mills dispersal. 
  The initial Bondi mass is $\MBi = 0.63$.
  At \scri (solid lines) the fields decay as $F \sim t^{-2}$ and 
  $D_F \sim t^{-1}$ at late times.
  At the origin (dashed lines), $F \sim D_F \sim t^{-4}$.
}
\end{figure}

Consider now the gravitational sector of Yang-Mills.
As for the scalar field, we begin with an evolution that disperses to flat
space (figure \ref{f:ym1disp}).
We find that the potential $F$ decays as $F \sim t^{-2}$ at \scri
and $F \sim t^{-4}$ at the origin.
These exponents agree with \cite{Puerrer2009}.
Let us also evaluate the electric field $D_F$. 
Whereas this decays at the same rate as $F$ at finite radius, we observe a 
slower decay $D_F \sim t^{-1}$ at \scri.
This may seem surprising at first but can be explained as follows.
Consider the evolution equation \eref{e:dtF} for $F$,
\begin{equation}
  \label{e:dtFgrav}
  \dot F = rX F' + 2XF - \tN D_F.
\end{equation}
Notice the radial derivative $F'$ on the right-hand side.
Suppose $F'$ decayed at the same rate as $F$ at \scri.
Then, by continuity, $F$ would decay at the same rate in a \emph{neighbourhood}
of \scri.
But we know from previous work \cite{Zenginoglu2008b,Puerrer2009}---and our
numerical evolutions suggest this too---that $F$ decays faster at any point 
away from \scri than at \scri itself.
Hence by contradiction $F'$ must decay at a slower rate at \scri than $F$.
From \eref{e:dtFgrav} we deduce that $D_F$ must also decay at that slower rate.
(This argument cannot determine the precise decay rate of $D_F$ at \scri, 
it merely shows that it must decay more slowly than $F$.)

\begin{figure}
\centerline{\includegraphics[scale=0.24]{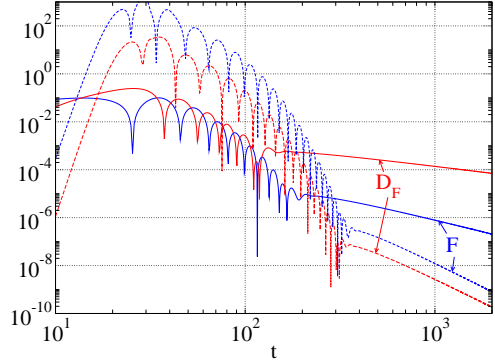}}
\caption{\label{f:ym1accr} 
  Gravitational-sector Yang-Mills accretion. 
  The initial Bondi and apparent horizon masses are $\MBi = 1.49$ and $\MAHi=1$,
  and the final masses are $\MBf = \MAHf = 1.45$.
  At \scri (solid lines) the fields decay as $F \sim t^{-2}$ and 
  $D_F \sim t^{-1}$ at late times.
  At the horizon (dashed lines), $F \sim D_F \sim t^{-4}$.
}
\end{figure}

Next we consider accretion onto a black hole (figure \ref{f:ym1accr}).
The decay exponents found in this case are the same as in the dispersing 
evolution and agree with the test field 
approximation \cite{Zenginoglu2008b}.

\begin{figure}
\centerline{\includegraphics[scale=0.24]{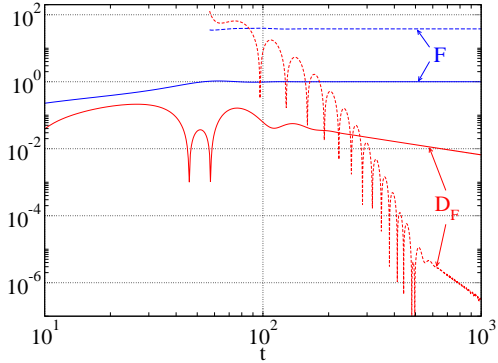}}
\caption{\label{f:ym1coll} 
  Gravitational-sector Yang-Mills collapse.
  The initial Bondi mass is $\MBi = 3.04$ and the final Bondi and apparent 
  horizon masses are $\MBf = \MAHf = 2.49$.
  The final coordinate location of the apparent horizon is $\rAH = 0.163$.
  At \scri (solid lines) $F \rightarrow -1$ and $D_F \sim t^{-1}$ at late times.
  At the horizon (after it forms, dashed lines), 
  $F \rightarrow -37.9 \approx -\rAH^{-2}$ and $D_F \sim t^{-4}$.
}
\end{figure}

Finally we turn to collapse (figure \ref{f:ym1coll}).
While the electric field decays at the same rates as above, the behaviour of 
the potential $F$ is different.
At late times this approaches the static solution $F=2/(gr^2)$ (recall we set 
$g=-2$).
We shall see in section \ref{s:ymvac} that this is another vacuum solution
of the Yang-Mills equations (in addition to $F=0$).

%%%%%%%%%%%%%%%%%%%%%%%%%%%%%%%%%%%%%%%%%%%%%%%%%%%%%%%%%%%%%%%%%%%%%%%%%%%%%%%%

\subsection{Sphaleron-sector Yang-Mills evolutions}

\begin{figure}
\includegraphics[width=0.49\textwidth]{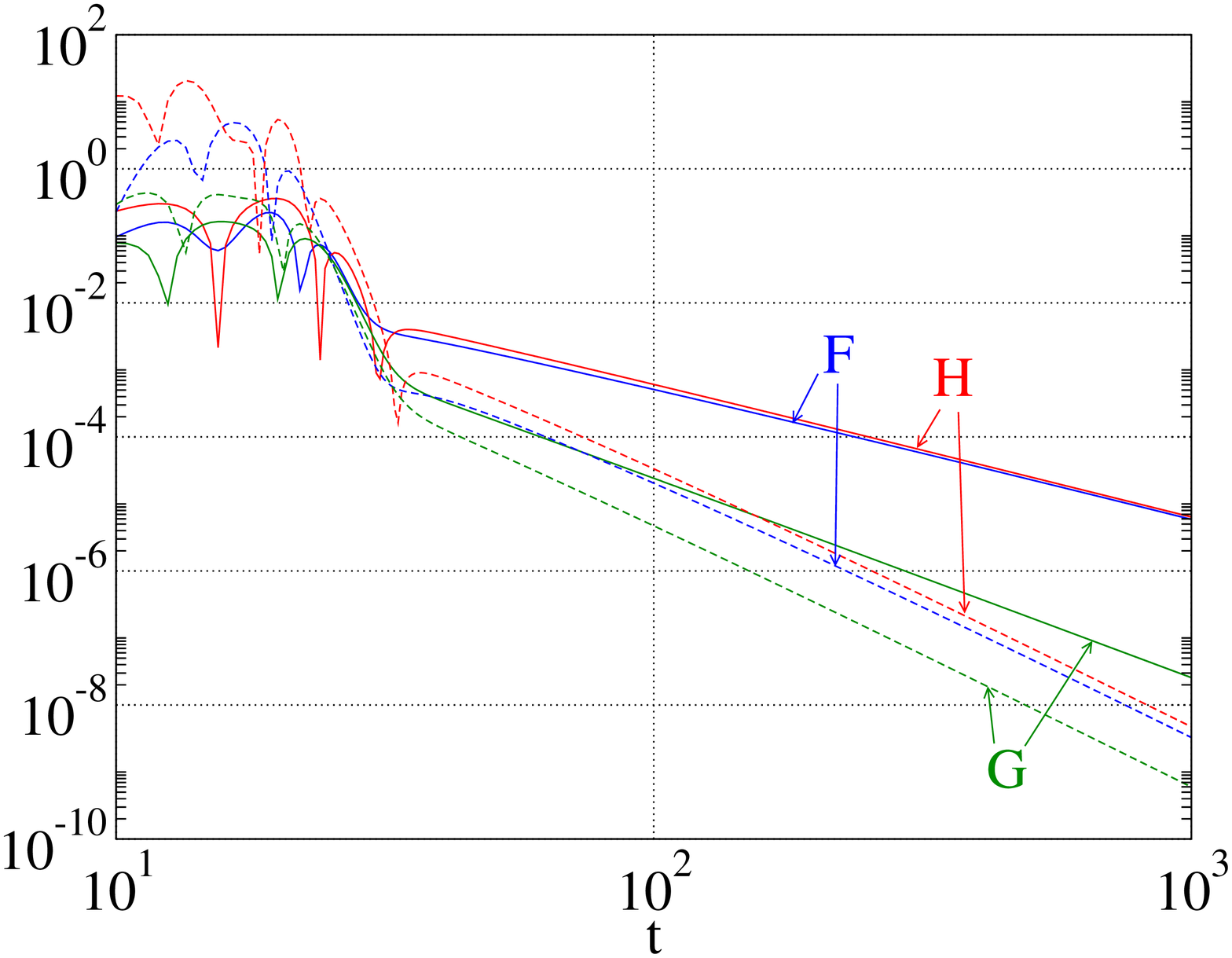}\hfill
\includegraphics[width=0.49\textwidth]{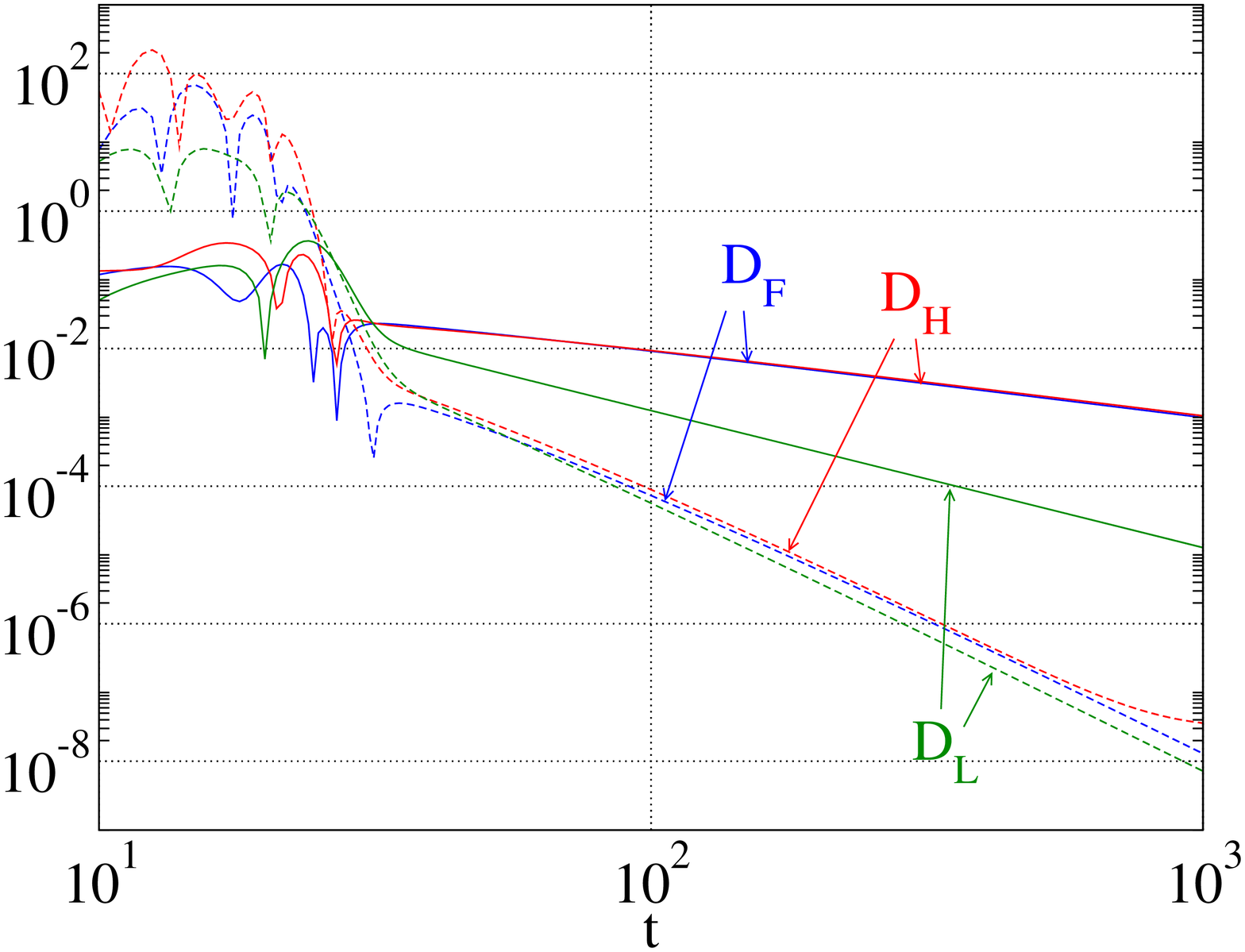}
\caption{\label{f:ym2disp} 
  Sphaleron-sector Yang-Mills dispersal. 
  The initial Bondi mass is $\MBi = 0.75$.
  At \scri (solid lines) the fields decay as $F \sim H \sim t^{-2}$,
  $G \sim t^{-3}$, $D_F \sim D_H \sim t^{-1}$ and $D_L \sim t^{-2}$. 
  At the origin (dashed lines), $F \sim H \sim G \sim D_F \sim D_H \sim D_L 
  \sim t^{-4}$.
}
\end{figure}

In the last set of evolutions we consider initial data of type (iii) (section
\ref{s:ini}) that include the sphaleron sector of Yang-Mills.
Figure \ref{f:ym2disp} shows a dispersing evolution.
All fields decay.
The potentials $F$ and $H$ decay at the same rate as in the gravitational 
sector, i.e., $F \sim H \sim t^{-2}$ at \scri and $F \sim H \sim t^{-4}$ at the
origin.
The ``gauge potential'' $G$ decays as $G \sim t^{-3}$ at \scri and 
$G \sim t^{-4}$ at the origin.
The electric field components $D_F$ and $D_H$ also decay at the same rates as
in the gravitational sector, i.e., $D_F \sim D_H \sim t^{-1}$ at \scri and 
$D_F \sim D_H \sim t^{-4}$ at the origin.
The component $D_L$ has a different decay at \scri, $D_L \sim t^{-2}$, 
while $D_L \sim t^{-4}$ at the origin as for the other components.

\begin{figure}
\includegraphics[width=0.49\textwidth]{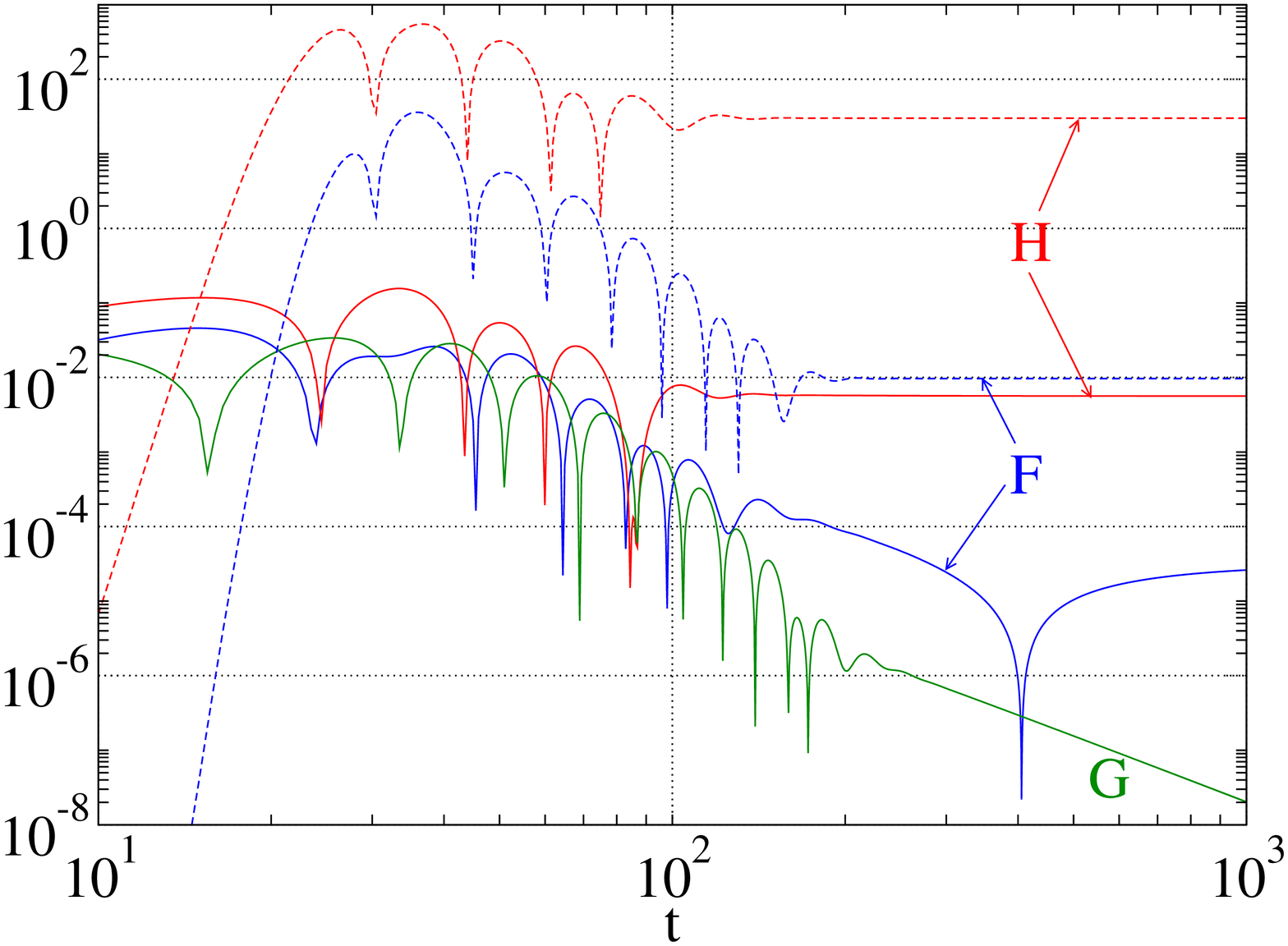}\hfill
\includegraphics[width=0.49\textwidth]{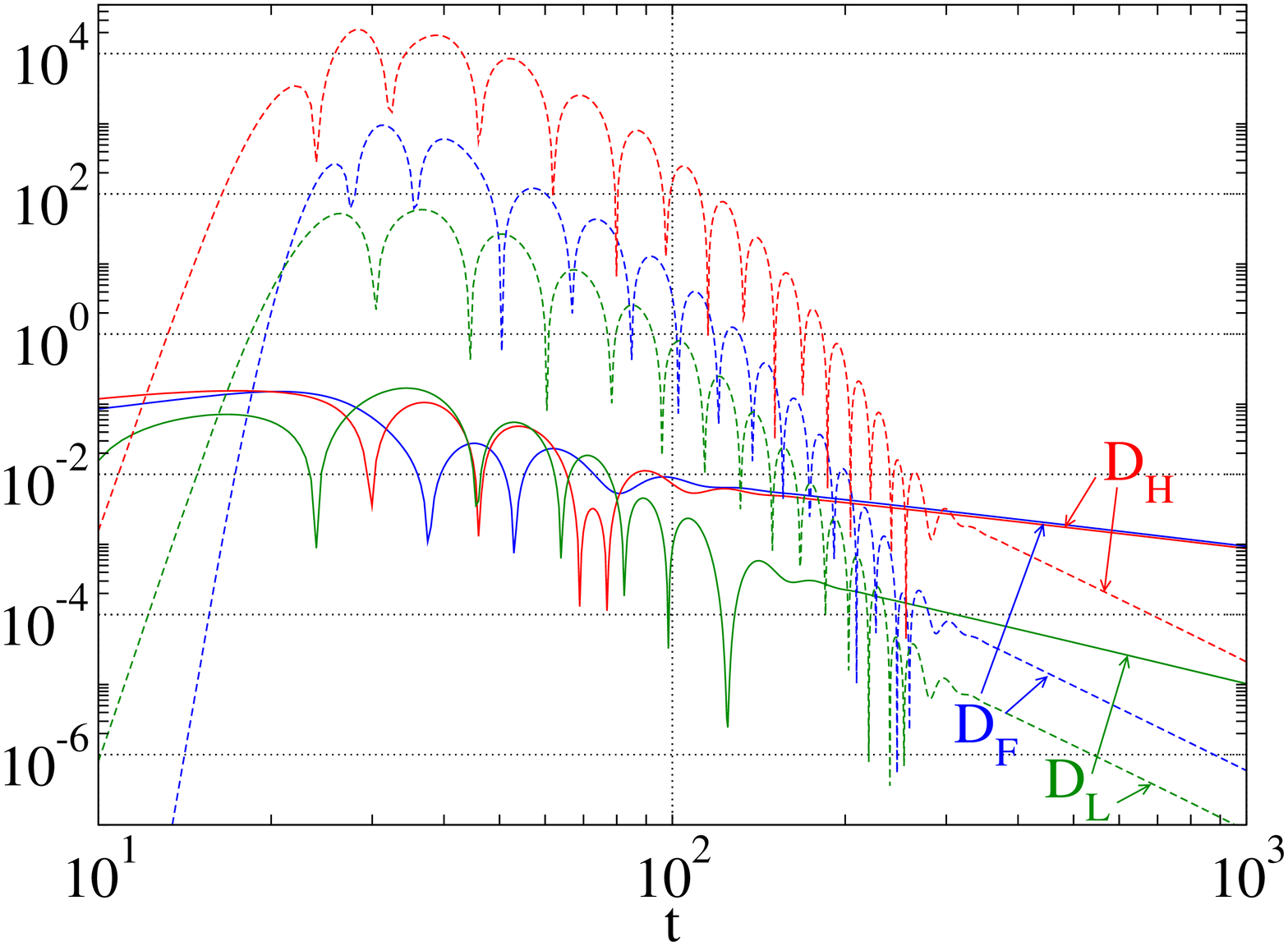}
\caption{\label{f:ym2accr} 
  Sphaleron-sector Yang-Mills accretion. 
  The initial Bondi and apparent horizon masses are $\MBi = 1.61$ and 
  $\MAHi = 1$, and the final masses are $\MBf = \MAHf = 2.49$.
  At \scri (solid lines), $F$ and $H$ approach constants, $G \sim t^{-3}$,
  $D_F \sim D_H \sim t^{-1}$ and $D_L \sim t^{-2}$.
  At the horizon (dashed lines), $F$ and $H$ approach constants,
  $G$ is frozen to its initial value $G=0$, and 
  $D_F \sim D_H \sim D_L \sim t^{-4}$.
}
\end{figure}

Already in the accretion evolution (figure \ref{f:ym2accr}), a different 
behaviour is seen for the connection components.
Now $F$ and $H$ approach a nonzero static solution at late times.
The gauge potential $G$ is frozen to its initial value at the excision
boundary, which for the initial data we choose is $G=0$ to within numerical 
roundoff.
At \scri, $G \sim t^{-3}$ as previously in the dispersing evolution.
The decay of the eletric field is also the same as in the dispersing evolution.

\begin{figure}
\includegraphics[width=0.49\textwidth]{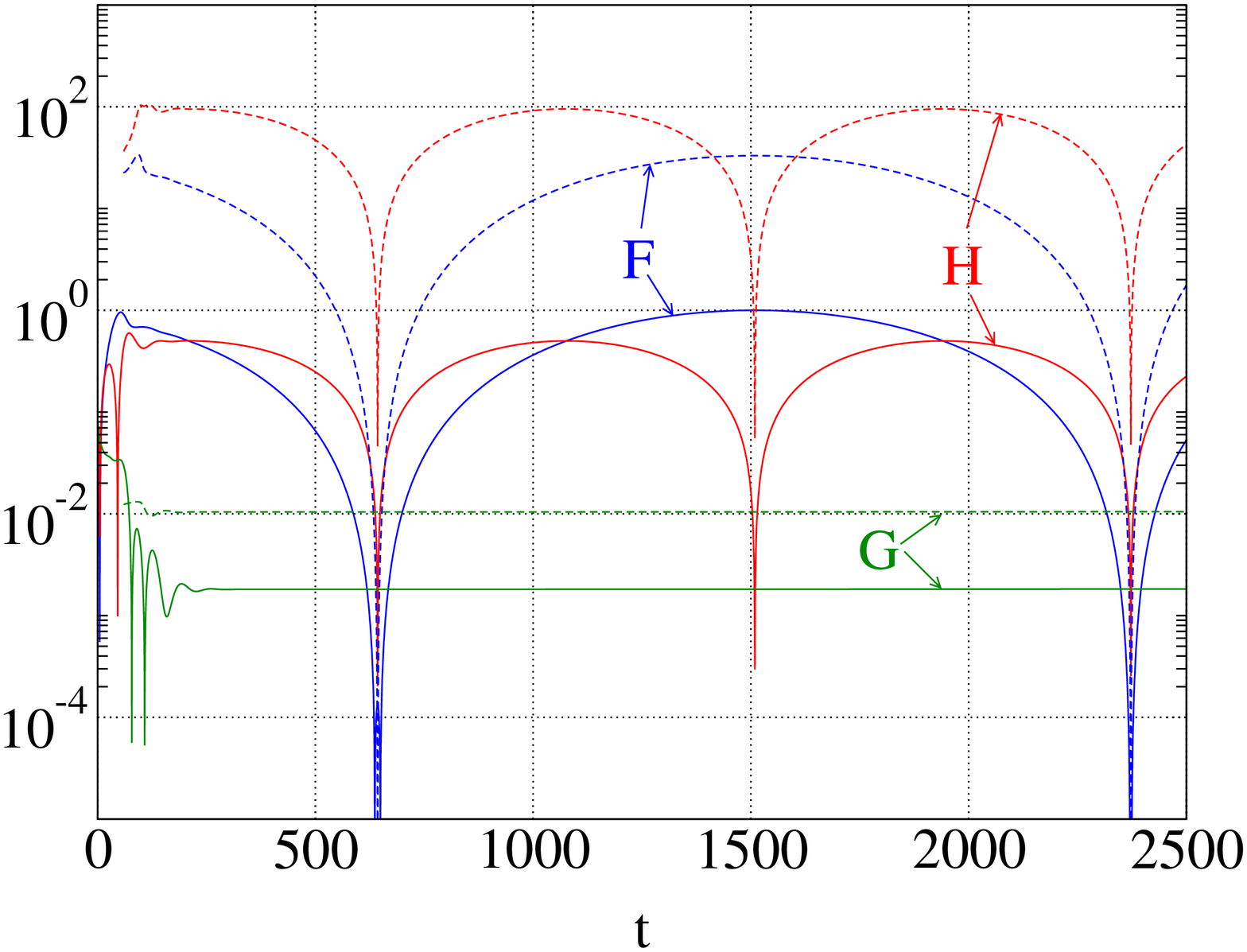}\hfill
\includegraphics[width=0.49\textwidth]{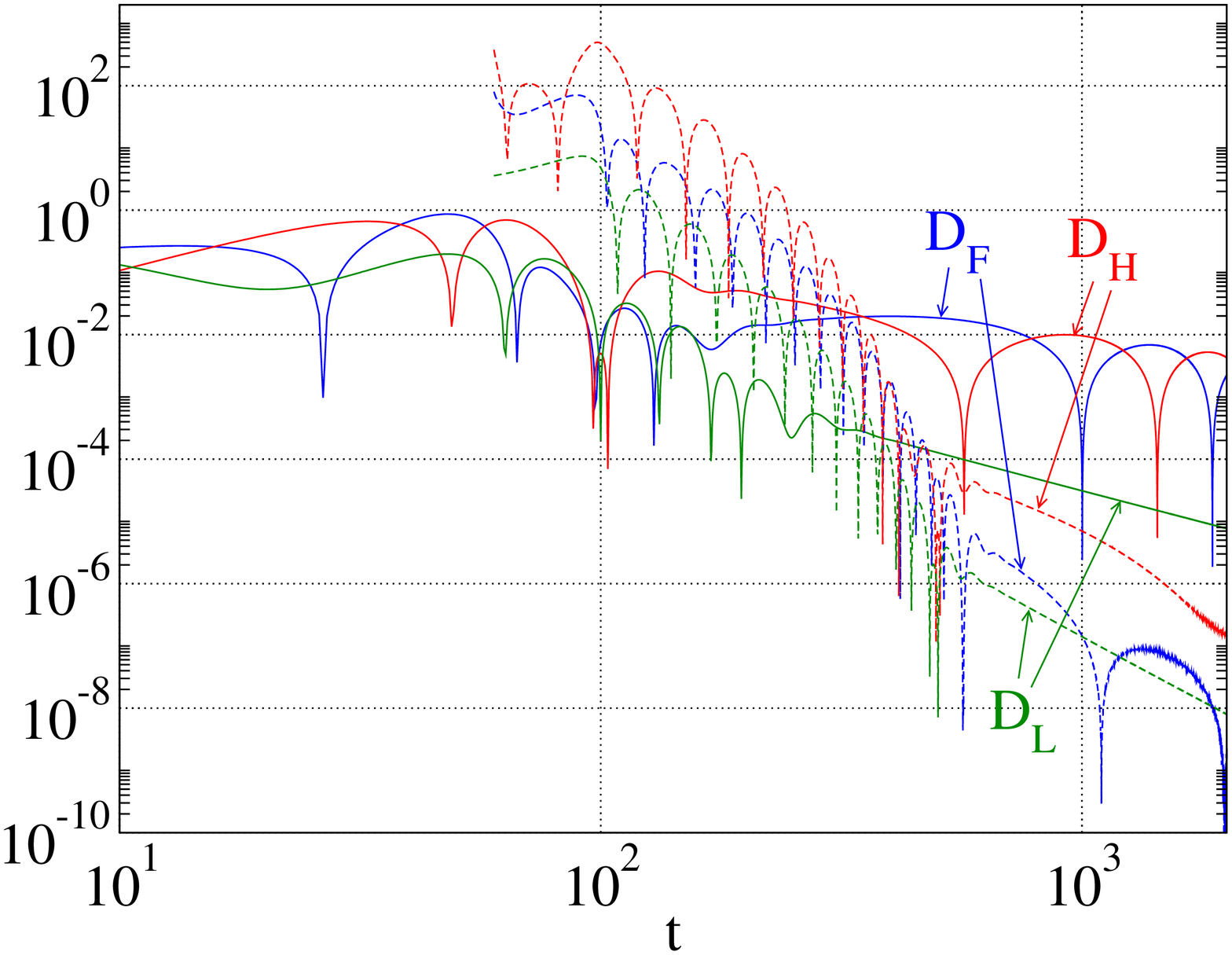}\medskip\\
\centerline{\includegraphics[width=0.5\textwidth]{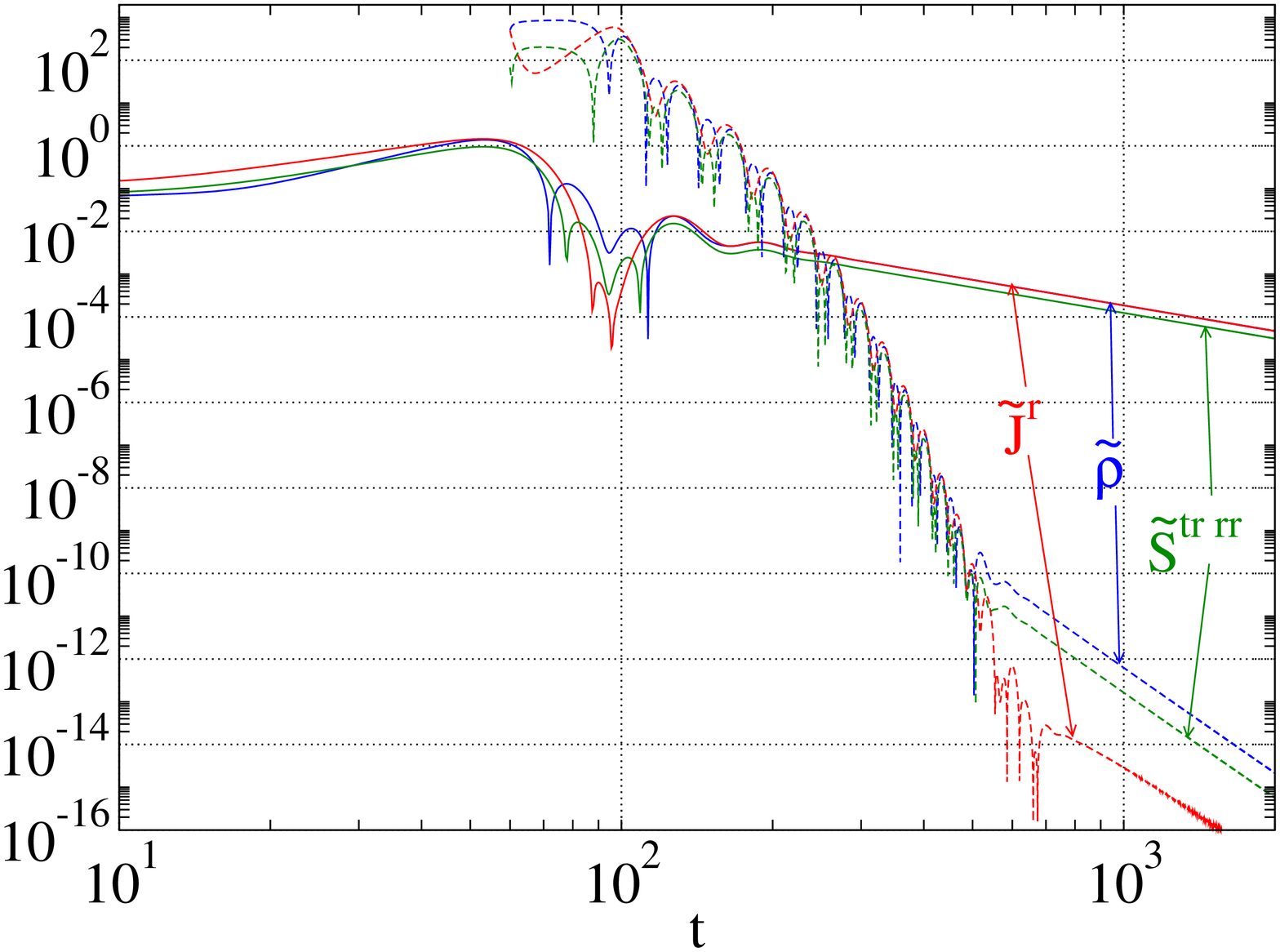}}
\caption{\label{f:ym2coll}  
  Sphaleron-sector Yang-Mills collapse. 
  The initial Bondi mass is $\MBi = 3.51$ and the final Bondi and apparent 
  horizon masses are $\MBf = \MAHf = 2.67$.
  As before, solid lines refer to quantities observed at \scri
  and dashed lines to quantities observed at the horizon after it forms.
  The top left panel shows that $G$ becomes asymptotically time independent, 
  $G \rightarrow G_0 = 0.00183$ at \scri, 
  and $F$ and $H$ perform harmonic oscillations with period 
  $T = 1717 \approx 2\pi/|gG_0|$. 
  In the top right panel, $D_F$ and $D_H$ decay in damped harmonic
  oscillations, $D_L \sim t^{-2}$ at \scri and $D_L \sim t^{-4}$ 
  at the horizon. 
  The bottom panel demonstrates that all independent components 
  $\tilde J^r$, $\tilde \rho$ and $\tilde S^{\tr rr}$ of the energy-momentum
  tensor decay as $t^{-2}$ at \scri and $t^{-8}$ at the horizon.
}
\end{figure}

Finally we study sphaleron-sector collapse (figure \ref{f:ym2coll}).
Recall that we freeze the value of $G$ at the excision boundary once the
apparent horizon forms.
This results in the entire function $G$ becoming time independent (but
nonzero).
Surprisingly however, the potentials $F$ and $H$ do not become time independent
but show a sinusoidal time dependence with a period close to $T = 2\pi/|g G_0|$,
where $G_0$ is the asymptotic value of $G$ at \scri.
This peculiar behaviour will be explained in section \ref{s:ymvac}.
A first hint at what is happening can be found by evaluating the components
of the energy-momentum tensor. 
Figure \ref{f:ym2coll} demonstrates that all of these decay at the rate
$t^{-2}$ at \scri and $t^{-8}$ at the horizon, in accordance with the electric 
field decaying as $t^{-1}$ at \scri and $t^{-4}$ at the horizon 
(recall that the energy-momentum tensor is quadratic in the field strength 
tensor).
Hence a vacuum solution is approached.

%%%%%%%%%%%%%%%%%%%%%%%%%%%%%%%%%%%%%%%%%%%%%%%%%%%%%%%%%%%%%%%%%%%%%%%%%%%%%%%%

\subsection{The Yang-Mills vacuum}
\label{s:ymvac}

The findings of the previous subsection suggest that we need to obtain a 
better understanding of the vacuum solutions to the Yang-Mills equations.
Let us therefore take the field strength tensor to vanish, i.e.,
$D_F = D_H = D_L = B_F = B_H = B_L = 0$.
Equation \eref{e:BF} for the magnetic field implies 
\begin{equation}
  \label{e:Hvac}
  rH' + 3H = 0 \Rightarrow H = H_0(t) r^{-3}.
\end{equation}
The linear combination $r^2 \times$ \eref{e:BH} $-$ \eref{e:BK} yields
\begin{equation}
  \label{e:Fvac}
  rF' + 2F = 0 \Rightarrow F = F_0(t) r^{-2}.
\end{equation}
The radial gauge condition \eref{e:YMradial} implies
\begin{equation}
  \label{e:Gvac}
  rG' + G = 0 \Rightarrow G = G_0(t) r^{-1}.
\end{equation}
Since we freeze $G$ at the inner boundary after excision, we are only interested
in the case where $G_0$ is a constant.
Substituting \eref{e:Hvac}--\eref{e:Gvac} into the evolution equations 
\eref{e:dtF} and \eref{e:dtH} for $F$ and $H$, we obtain
\begin{equation}
  \label{e:odesys}
  \dot F_0 = -g G_0 H_0, \qquad
  \dot H_0 = -G_0 + g G_0 F_0.
\end{equation}
The general solution to this pair of ODEs is
\begin{equation}
  \label{e:YMgaugeosc}
  F_0 = g^{-1} + \alpha \sin(gG_0t + \phi), \qquad
  H_0 = -\alpha \cos (gG_0 t + \phi).
\end{equation}
The constant $\alpha$ is not arbitrary, however.
In deriving \eref{e:odesys} we used only one particular linear
combination of \eref{e:BH} and \eref{e:BK}.
Substituting the solution \eref{e:YMgaugeosc} in either \eref{e:BH} or 
\eref{e:BK} fixes $\alpha = g^{-1}$.
The form of the solution \eref{e:YMgaugeosc} agrees well with the numerical 
results in figure \ref{f:ym2coll}.
We see that it is $G_0$ (the value of $G$ at \scri) that determines
the period of the oscillations.
If $G$ is zero as in the final state of the accretion evolution,
\eref{e:YMgaugeosc} implies that $F_0$ and $H_0$ are constants related by
\begin{equation}
  (1-gF_0)^2 + (gH_0)^2 = 1.
\end{equation}
If in addition $H_0=0$ as in the gravitational-sector evolutions,
this implies $F_0 = 0$ or $F_0 = 2/g$, i.e., by \eref{e:Fvac},
\begin{equation}
  F = 0 \; \mathrm{or} \; F = \frac{2}{gr^2}.
\end{equation}
In the Abelian-gauge version \eref{e:abeliangauge} of the gravitational-sector 
Yang-Mills ansatz, these two copies of the Yang-Mills vacuum correspond to 
$w=\pm 1$.

%%%%%%%%%%%%%%%%%%%%%%%%%%%%%%%%%%%%%%%%%%%%%%%%%%%%%%%%%%%%%%%%%%%%%%%%%%%%%%%
%%%%%%%%%%%%%%%%%%%%%%%%%%%%%%%%%%%%%%%%%%%%%%%%%%%%%%%%%%%%%%%%%%%%%%%%%%%%%%%

\section{Conclusions}
\label{s:concl}

The purpose of this paper was to show how matter can be included in a 
hyperboloidal evolution scheme for the Einstein equations, and to implement 
such a scheme numerically in order to study power-law tails of matter fields 
up to future null infinity.

We assumed that the energy-momentum tensor is tracefree, for then the 
energy-momentum conservation equations are conformally invariant and thus
regular at null infinity.
A large class of radiative matter models are included under this assumption.
How to deal with non-tracefree matter in the conformal setting remains an 
open question.
Of course, if the matter is such that it remains bounded away from \scri during 
the evolution then one can live with matter field equations that are singular 
at \scri as they never need to be evaluated there.

We worked with a constrained ADM formulation of the Einstein equations on CMC
slices \cite{Moncrief2009}. 
In that paper we showed (in the vacuum case) how the formally singular terms 
in the evolution equations at \scri can in fact be evaluated in a regular way. 
In the present paper we showed that our analysis is 
unaffected by the addition of matter (section \ref{s:scrireg}).

Two matter models were studied in detail, namely a conformally coupled scalar 
field and a Yang-Mills field.
In both cases the energy-momentum tensor is tracefree and the matter equations 
of motion are conformally invariant.
The use of conformal coupling instead of minimal coupling for the scalar field
is essential here.
We first derived the matter evolution equations and the source terms appearing
in the Einstein equations without any symmetry assumptions.
Subsequently we reduced the Einstein-matter systems to spherical symmetry
in isotropic coordinates.
We worked with the most general ansatz \eref{e:spherym} for the spherically 
symmetric Yang-Mills connection \cite{Witten1977,Gu1981,SarbachPhD}; 
this is more general than the purely magnetic or gravitational-sector 
ansatz that is often used.

Our motivation to study spherically symmetric Einstein-matter evolutions on
hyperboloidal slices arose partially from an earlier numerical study by
one of the authors \cite{Rinne2010} that considered vacuum axisymmetric
spacetimes.
Whereas long-term stable evolutions of perturbed black holes could be achieved
and the quasi-normal mode radiation correctly reproduced, we were unable to
resolve the power-law tail of the gravitational field expected at late times.
In the current spherically symmetric study a much higher numerical resolution
could be used so that the tails could indeed be resolved.
Otherwise the numerical method based on fourth-order finite differences 
is very similar to the one used in \cite{Rinne2010}.
One difference is that we are now able to handle a regular centre as well
so that gravitational collapse can be studied; in \cite{Rinne2010} we only
considered black hole evolutions with excised interior.

A general feature of tails in hyperboloidal evolutions is that the fields
decay at a slower rate at \scri than at finite radius.
This means that the solution becomes increasingly peaked at \scri, akin to
the formation of a boundary layer.
This is very challenging numerically---at a fixed numerical resolution the 
code is ultimately unable to resolve the solution.
A possible direction for future development would be to introduce a 
time-dependent radial mapping so that an increasingly higher resolution 
can be provided near \scri as the evolution proceeds.
A similar adaptivity will be needed in order to study the formation of 
very small black holes in critical collapse.
The current code already contains an (albeit time-independent) radial map 
in order to better resolve the steep gradients near the horizon of a black hole.

For scalar field matter our results are consistent with evolutions of the
Einstein-scalar field system in Bondi coordinates \cite{Puerrer2005}.
Our study goes further though because unlike Bondi coordinates,
the coordinates we use can penetrate black hole horizons.
Both for an evolution that starts out with a perturbed black hole and for
an evolution that forms a black hole from regular initial data, we find the 
same decay exponents at \scri and at finite radius as in the dispersing case,
which agree with results obtained in the test field 
approximation \cite{Zenginoglu2008b}.

For evolutions of the gravitational sector of Yang-Mills, the decay rates 
we found for the Yang-Mills connection in an evolution that disperses to 
flat space agree with evolutions in Bondi coordinates \cite{Puerrer2009}.
For a Schwarzschild black hole with a Yang-Mills perturbation we obtain the
same decay rates, which in turn agree with the test field
approximation \cite{Zenginoglu2008b}.
Somewhat different behaviour is seen in an evolution that collapses to a black
hole from regular initial data.
Here the Yang-Mills connection approaches a static solution that corresponds to
another copy of the Yang-Mills vacuum.

A fact that does not seem to have been noted before is that in all three cases 
(dispersal, accretion and collapse), the electric field
has a slower decay at \scri than the connection, $D_F \sim t^{-1}$ as opposed
to $F \sim t^{-2}$. 
At finite radius the decay rates agree, $F \sim D_F \sim t^{-4}$.
Since the electric field is the physically measurable quantity, it seems 
more natural to consider its decay rather than that of the connection.

As far as we know we presented the first dynamical numerical evolutions that
include the sphaleron sector of Yang-Mills, i.e., the fully general ansatz 
\eref{e:spherym} for the spherically symmetric Yang-Mills connection.
In an evolution that disperses to Minkowski space, both potentials $F$ and $H$
appearing in the ansatz for the connection decay at the same rates as for the
gravitational-sector evolutions.
For an accretion evolution, however, both potentials approach nonzero static
solutions, and in a collapse evolution they become sinusoidal functions of time.
Nevertheless, in all cases a vacuum solution is approached.
We were able to explain this behaviour by deriving the general vacuum solution
to the Yang-Mills equations in our setting.

The nontrivial structure of the Yang-Mills vacuum gives rise to interesting
threshold behaviour \cite{Bizon2010}.
The Einstein-Yang-Mills system exhibits remarkably rich 
dynamics \cite{Choptuik1996,Choptuik1999} due to the existence of nontrivial 
asymptotically flat static solutions \cite{Bartnik1988,Bizon1990}.
We hope to investigate some of these phenomena further using our hyperboloidal
evolution code.

%%%%%%%%%%%%%%%%%%%%%%%%%%%%%%%%%%%%%%%%%%%%%%%%%%%%%%%%%%%%%%%%%%%%%%%%%%%%%%%
%%%%%%%%%%%%%%%%%%%%%%%%%%%%%%%%%%%%%%%%%%%%%%%%%%%%%%%%%%%%%%%%%%%%%%%%%%%%%%%

\section*{Acknowledgments}

The authors would like to thank Lars Andersson, Piotr Bizo\'n, Helmut Friedrich,
Michael P\"urrer, Istv\'an R\'acz, Olivier Sarbach, Christian Schell and 
An{\i}l Zengino\u{g}lu for helpful discussions.
O.R. gratefully acknowledges support from the German Research Foundation (DFG)
through a Heisenberg Fellowship and research grant RI 2246/2.
V.M. was supported by NSF grant PHY-0963869 to Yale University.
V.M. is grateful to the Albert Einstein Institute (Potsdam), the Erwin 
Schr\"odinger Institute and the University of Vienna for hospitality and 
support during the course of some of this work.

%%%%%%%%%%%%%%%%%%%%%%%%%%%%%%%%%%%%%%%%%%%%%%%%%%%%%%%%%%%%%%%%%%%%%%%%%%%%%%%
%%%%%%%%%%%%%%%%%%%%%%%%%%%%%%%%%%%%%%%%%%%%%%%%%%%%%%%%%%%%%%%%%%%%%%%%%%%%%%%

\appendix

\section{Conformal transformations and $3+1$ decompositions}
\label{s:app}

In this appendix we collect a few useful identities used throughout the paper.

\subsection{Conformal transformation of the connection and curvature}

Consider a conformal transformation of the metric,
\begin{equation}
  \gamma_{ab} = \Omega^2 g_{ab}.
\end{equation}
The following applies to any dimension $n$ and any signature (e.g. the 
spacetime metric or the induced spatial metric on the $t=\const$ slices), 
and hence we use general indices $a,b,\ldots$ to indicate this. 

Let $\Gamma^a{}_{bc}$ denote the Christoffel symbols of the metric connection
$\nabla$ of $g$ and $\tGamma^a{}_{bc}$ the Christoffel symbols of the 
metric connection $\tnabla$ of $\gamma$.
They are related by
\begin{equation}
  \Gamma^a{}_{bc} = \tGamma^a{}_{bc} - \Omega^{-1} \left(
    \delta^a_b \Omega_{,c} + \delta^a_c \Omega_{,b}
    - \gamma_{bc}\gamma^{ad}\tnabla_d \Omega \right).
\end{equation}
The Ricci tensor transforms as
\begin{eqnarray}
  R_{ab} &=& \tR_{ab} + (n-2) \Omega^{-1}\tnabla_a \tnabla_b \Omega\nonumber\\
  &&+ \gamma_{ab} \gamma^{cd} \left[\Omega^{-1} \tnabla_c\tnabla_d\Omega
    - (n-1)\Omega^{-2}\Omega_{,c}\Omega_{,d} \right]
\end{eqnarray}
and the Ricci scalar as
\begin{equation}
  \label{e:Rtrafo}
  R = \Omega^2 \tR + 2(n-1)\Omega\gamma^{ab}\tnabla_a\tnabla_b\Omega
  - n(n-1)\gamma^{ab}\Omega_{,a}\Omega_{,b}.
\end{equation}

%%%%%%%%%%%%%%%%%%%%%%%%%%%%%%%%%%%%%%%%%%%%%%%%%%%%%%%%%%%%%%%%%%%%%%%%%%%%%%%

\subsection{Physical and conformal extrinsic curvature}

The physical extrinsic curvature $K_{ij}$ and the unphysical extrinsic
curvature $C_{ij}$ are defined as 
\begin{equation}
  K_{ij} = -\half \Lie_n g_{ij},\qquad
  C_{ij} = -\half \Lie_{\tilde n} \gamma_{ij}.
\end{equation}
By translating the right-hand side of the first equation into conformal 
language and comparing with the second, we find for the traceless part
\begin{equation}
  \label{e:c_k_tr}
  C_{ij}^{\tr} = \Omega K_{ij}^{\tr} =-\mu_\gamma^{-1} \gamma_{ik}\gamma_{jl}
  \pi^{\tr kl}. 
\end{equation}
Taking instead the trace, we recover equation \eref{e:dtOmega},
\begin{equation}
  \Lie_{\tn} \Omega = -\third (K + \Omega C),
\end{equation}
where $C \equiv \gamma^{ij}C_{ij}$ and we recall our convention 
$K \equiv -g^{ij}K_{ij} = \const > 0$.

%%%%%%%%%%%%%%%%%%%%%%%%%%%%%%%%%%%%%%%%%%%%%%%%%%%%%%%%%%%%%%%%%%%%%%%%%%%%%%%

\subsection{$3+1$ decomposition of the scalar Hessian}
\label{s:scalhess}

The projections of the covariant Hessian $\ftnabla_\mu \ftnabla_\nu \tphi$ 
of a conformal scalar field $\tphi$ can be written in $3+1$ form as
\begin{eqnarray}
  n^\mu n^\nu  \ftnabla_\mu \ftnabla_\nu \tphi 
  &=& \Lie_{\tn}^2 \tphi - \tN^{-1} \gamma^{ij} \tN_{,i} \tphi_{,j},\\
  \gamma^{i\mu} n^\nu \ftnabla_\mu \ftnabla_\nu \tphi
  &=& \partial^i \Lie_{\tn}\tphi + C^{ij}\tphi_{,j},\\
  \gamma_i{}^\mu \gamma_j{}^\nu \ftnabla_\mu \ftnabla_\nu \tphi
  &=& \tnabla_i \tnabla_j \tphi + C_{ij} \Lie_{\tn} \tphi.
\end{eqnarray}

%%%%%%%%%%%%%%%%%%%%%%%%%%%%%%%%%%%%%%%%%%%%%%%%%%%%%%%%%%%%%%%%%%%%%%%%%%%%%%%

\subsection{$3+1$ decomposition of the conformal spacetime Ricci 
  tensor}
\label{s:confricc}

One of the equations of a $3+1$ decomposition in the conformal spacetime is
\begin{equation}
  \label{e:Cijevol}
  \fl \gamma_i{}^\mu \gamma_j{}^\nu \ftR_{\mu\nu} 
  = - \Lie_{\tilde n} C_{ij} 
  - 2 \gamma^{kl} C_{ik}C_{jl} + C_{ij} C - \tN^{-1}\tnabla_i\tnabla_j\tN 
  + \tR_{ij} .
\end{equation}
Separating the trace and traceless part of $C_{ij}$ and using \eref{e:c_k_tr},
we can write the traceless part of \eref{e:Cijevol} as
\begin{eqnarray}
  \label{e:Cijevol2}
  \fl [ \gamma^{i\mu} \gamma^{j\nu} \ftR_{\mu\nu} ]^{\tr} 
  = \mu_\gamma^{-1} \Lie_{\tilde n} \pi^{\tr ij} 
  + 2\mu_\gamma^{-2} \gamma_{kl} \pi^{\tr ik}\pi^{\tr jl}
  - \tfrac{2}{3} C \mu_\gamma^{-1} \pi^{\tr ij} \nonumber\\
  - \tN^{-1} [ \tnabla^i \tnabla^j \tN ]^{\tr}
  + \tR^{\tr ij} .  
\end{eqnarray}
Note the similarity of \eref{e:Cijevol2} with \eref{e:dtpi};
formally it can be obtained from that equation by setting $\Omega\equiv 1$,
replacing $K$ with $-C$ and noting that the source term now refers to the 
conformal geometry.
Taking instead the trace of \eref{e:Cijevol} we find
\begin{equation}
  \label{e:Cevol}
   \gamma^{\mu\nu} \ftR_{\mu\nu}= - \tN^{-1}\tnabla^i\tnabla_i\tN + \tR + C^2
  - \Lie_{\tilde n} C .
\end{equation}

We also have the Gauss-Codazzi equation analogous to the Hamiltonian 
constraint \eref{e:hamcons},
\begin{eqnarray}
  \label{e:conf_GC}
  \fl \tn^\mu\tn^\nu\ftR_{\mu\nu} + \gamma^{\mu\nu}\ftR_{\mu\nu}
  = 2 \tn^\mu\tn^\nu \four{\tilde G}_{\mu\nu} \nonumber\\
  = \tR + C^2 - \gamma^{ik}\gamma^{jl}C_{ij}C_{kl} 
  = \tR + \tfrac{2}{3} C^2 
  - \mu_\gamma^{-2} \gamma_{ik}\gamma_{jl}\pi^{\tr ij} \pi^{\tr kl}.
\end{eqnarray}
Taking linear combinations of \eref{e:Cevol} and \eref{e:conf_GC}, we obtain
\begin{eqnarray}
  \fl \tn^\mu\tn^\nu\ftR_{\mu\nu} = \tN^{-1}\tnabla^i\tnabla_i\tN - \third C^2
  - \mu_\gamma^{-2} \gamma_{ik}\gamma_{jl}\pi^{\tr ij}\pi^{\tr kl} + \Lie_{\tn} C, \\
  \label{e:ftR}
  \fl \ftR = \gamma^{\mu\nu}\ftR_{\mu\nu} - \tn^\mu\tn^\nu\ftR_{\mu\nu} 
  \nonumber\\
  = -2 \tN^{-1}\tnabla^i\tnabla_i\tN + \tR + \tfrac{4}{3} C^2 
  + \mu_\gamma^{-2} \gamma_{ik}\gamma_{jl}\pi^{\tr ij}\pi^{\tr kl} - 2 \Lie_{\tn} C.
\end{eqnarray}

Finally, the Gauss-Codazzi equation analogous to the momentum 
constraint \eref{e:momcons} is
\begin{equation}
  \gamma^{i\mu} \tn^\nu \ftR_{\mu\nu} = \mu_\gamma^{-1} \tnabla_j \pi^{\tr ij}
  + \tfrac{2}{3} \gamma^{ij} C_{,j}.
\end{equation}

%%%%%%%%%%%%%%%%%%%%%%%%%%%%%%%%%%%%%%%%%%%%%%%%%%%%%%%%%%%%%%%%%%%%%%%%%%%%%%%
%%%%%%%%%%%%%%%%%%%%%%%%%%%%%%%%%%%%%%%%%%%%%%%%%%%%%%%%%%%%%%%%%%%%%%%%%%%%%%%

\section*{References}

%\bibliographystyle{oriop}
%\bibliography{bibtex/References}

\begin{thebibliography}{10}
\expandafter\ifx\csname url\endcsname\relax
  \def\url#1{{\tt #1}}\fi
\expandafter\ifx\csname urlprefix\endcsname\relax\def\urlprefix{\textit{URL}
  }\fi
\providecommand{\eprint}[2][]{\url{#2}}

\bibitem{SarbachLRR}
Sarbach O and Tiglio M 2012 Continuum and discrete initial-boundary value
  problems and {Einstein's} field equations {\em Living Rev.\ Relativity\/}
  {\bf 15}(9)

\bibitem{Penrose1965}
Penrose R 1965 Zero rest-mass fields including gravitation: Asymptotic
  behaviour {\em Proc.\ Royal Soc.\ London A\/} {\bf 284} 159--203

\bibitem{Moncrief2009}
Moncrief V and Rinne O 2009 Regularity of the {E}instein equations at future
  null infinity {\em Class.\ Quantum Grav.\/} {\bf 26} 125010
  %\url{http://dx.doi.org/10.1088/0264-9381/26/12/125010} \textit{Preprint}
  %\eprint{http://www.arxiv.org/abs/0811.4109} (One of the journal's
  %``Highlights of 2008 and 2009'')

\bibitem{Arnowitt1962}
Arnowitt R, Deser S and Misner C~W 1962 The dynamics of general relativity {\em
  Gravitation: an introduction to current research\/} ed Witten L (New York:
  Wiley) chap~7

\bibitem{Rinne2010}
Rinne O 2010 An axisymmetric evolution code for the {E}instein equations on
  hyperboloidal slices {\em Class.\ Quantum Grav.\/} {\bf 27} 035014
  %\url{http://dx.doi.org/10.1088/0264-9381/27/3/035014} \textit{Preprint}
  %\eprint{http://www.arxiv.org/abs/0910.0139}

\bibitem{Friedrich1983a}
Friedrich H 1983 Cauchy problems for the conformal vacuum field equations in
  general relativity {\em Commun.\ Math.\ Phys.\/} {\bf 91} 445--472

\bibitem{FrauendienerLRR}
Frauendiener J 2004 Conformal infinity {\em Living Rev.\ Relativity\/} {\bf
  7}(1)

\bibitem{Husa2002}
Husa S 2002 Problems and successes in the numerical approach to the conformal
  field equations {\em Lect.~Notes Phys.\/} {\bf 604} 239--260

\bibitem{Husa2003}
Husa S 2003 Numerical relativity with the conformal field equations {\em
  Lect.~Notes Phys.\/} {\bf 617} 159--192

\bibitem{Huebner1995}
H\"ubner P 1995 General relativistic scalar-field models and asymptotic
  flatness {\em Class.\ Quantum Grav.\/} {\bf 12} 791--808

\bibitem{Huebner1996}
H\"ubner P 1996 A method for calculating the structure of (singular) spacetimes
  in the large {\em Phys.\ Rev.\ D\/} {\bf 53} 701--721

\bibitem{Zenginoglu2008}
Zengino\u{g}lu A 2008 Hyperbolodial evolution with the {E}instein equations
  {\em Class.\ Quantum Grav.\/} {\bf 25} 195025

\bibitem{Bardeen2011}
Bardeen J~M, Sarbach O and Buchman L~T 2011 Tetrad formalism for numerical
  relativity on conformally compactified constant mean curvature hypersurfaces
  {\em Phys.\ Rev.\ D\/} {\bf 83} 104045

\bibitem{Buchman2009}
Buchman L~T, Pfeiffer H~P and Bardeen J~M 2009 Black hole initial data on
  hyperboloidal slices {\em Phys.\ Rev.\ D\/} {\bf 80} 084024

\bibitem{Bardeen2012}
Bardeen J~M and Buchman L~T 2012 {Bondi-Sachs} energy-momentum for the constant
  mean extrinsic curvature initial value problem {\em Phys.\ Rev.\ D\/} {\bf
  85} 064035

\bibitem{Price1972}
Price R~H 1972 Nonspherical perturbations of relativistic gravitational
  collapse. {I.} scalar and gravitational perturbations {\em Phys.\ Rev.\ D\/}
  {\bf 5} 2419--2439

\bibitem{Dafermos2005}
Dafermos M and Rodnianski I 2005 A proof of {Price's} law for the collapse of a
  self-gravitating scalar field {\em Inventiones Mathematicae\/} {\bf 162}
  381--457

\bibitem{Puerrer2005}
P\"urrer M, Husa S and Aichelburg P~C 2005 News from critical collapse: {B}ondi
  mass, tails, and quasinormal modes {\em Phys.\ Rev.\ D\/} {\bf 71} 104005

\bibitem{Leaver1986}
Leaver E~W 1986 Spectral decomposition of the perturbation response of the
  {Schwarzschild} geometry {\em Phys.\ Rev.\ D\/} {\bf 34} 384--408

\bibitem{Gundlach1994}
Gundlach C, Price R~H and Pullin J 1994 Late-time behavior of stellar collapse
  and explosions. {I.} linearized perturbations {\em Phys.\ Rev.\ D\/} {\bf 49}
  883--889

\bibitem{Gundlach1994a}
Gundlach C, Price R~H and Pullin J 1994 Late-time behavior of stellar collapse
  and explosions. {II.} nonlinear evolution {\em Phys.\ Rev.\ D\/} {\bf 49}
  890--899

\bibitem{Zenginoglu2008b}
Zengino\u{g}lu A 2008 A hyperboloidal study of tail decay rates for scalar and
  {Yang-Mills} fields {\em Class.\ Quantum Grav.\/} {\bf 25} 175013

\bibitem{Bizon2007}
Bizo\'n P, Chmaj T and Rostworowski A 2007 Late-time tails of a {Yang-Mills}
  field on {Minkowski} and {Schwarzschild} backgrounds {\em Class.\ Quantum
  Grav.\/} {\bf 24} F55--F63

\bibitem{Puerrer2009}
P\"urrer M and Aichelburg P~C 2009 Tails for the {Einstein-Yang-Mills} system
  {\em Class.\ Quantum Grav.\/} {\bf 26} 035004

\bibitem{Friedrich1991}
Friedrich H 1991 On the global existence and the asymptotic behavior of
  solutions to the {Einstein-Maxwell-Yang-Mills} equations {\em J.\ Diff.\
  Geom.\/} {\bf 34} 275--345

\bibitem{Luebbe2011a}
L\"ubbe C and Valiente~Kroon J~A 2013 A conformal approach for the analysis of
  the non-linear stability of pure radiation cosmologies {\em Annals Phys.\/}
  {\bf 328} 1--25

\bibitem{Winicour1988}
Winicour J 1988 Massive fields at null infinity {\em J.\ Math.\ Phys.\/} {\bf
  29} 2117--2121

\bibitem{York1979}
York Jr J~W 1979 Kinematics and dynamics of general relativity {\em Sources of
  gravitational radiation\/} ed Smarr L~L (Cambridge University Press) pp
  83--126

\bibitem{Andersson1992}
Andersson L, Chru\'sciel P~T and Friedrich H 1992 On the regularity of
  solutions to the {Y}amabe equation and the existence of smooth hyperboloidal
  initial data for {E}instein's field equations {\em Commun.\ Math.\ Phys.\/}
  {\bf 149} 587--612

\bibitem{Bekenstein1974}
Bekenstein J~D 1974 Exact solutions of {E}instein-conformal scalar equations
  {\em Ann.\ Phys.\/} {\bf 82} 535--547

\bibitem{Rinne2005}
Rinne O and Stewart J~M 2005 A strongly hyperbolic and regular reduction of
  {E}instein's equations for axisymmetric spacetimes {\em Class.\ Quantum
  Grav.\/} {\bf 22} 1143--1166
  %\url{http://dx.doi.org/10.1088/0264-9381/22/6/015} \textit{Preprint}
  %\eprint{http://www.arxiv.org/abs/gr-qc/0502037} (One of the journal's
  %``Highlights of 2004 and 2005'')

\bibitem{Ruiz2007a}
Ruiz M, Alcubierre M and N\'u\~nez D 2007 Regularization of spherical and
  axisymmetric evolution codes in numerical relativity {\em Gen.\ Relativ.\
  Gravit.\/} {\bf 40} 159--182

\bibitem{Witten1977}
Witten E 1977 Some exact multipseudoparticle solutions of classical
  {Yang-Mills} theory {\em Phys.\ Rev.\ Lett.\/} {\bf 38} 121--124

\bibitem{Gu1981}
Gu C and Hu H 1981 On the spherically symmetric gauge fields {\em Commun.\
  Math.\ Phys.\/} {\bf 79} 75--90

\bibitem{SarbachPhD}
Sarbach O 2000 {\em On the generalization of the {Regge-Wheeler} equation for
  self-gravitating matter fields\/} PhD thesis University of Zurich

\bibitem{Choptuik1999}
Choptuik M~W, Hirschmann E~W and Marsa R~L 1999 New critical behavior in
  {Einstein-Yang-Mills} collapse {\em Phys.\ Rev.\ D\/} {\bf 60} 124011

\bibitem{Choptuik1996}
Choptuik M~W, Chmaj T and Bizo\'n P 1996 Critical behavior in gravitational
  collapse of a {Yang-Mills} field {\em Phys.\ Rev.\ Lett.\/} {\bf 77} 424--427

\bibitem{Fodor2008}
Fodor G and R\'acz I 2008 Numerical investigation of highly excited magnetic
  monopoles in {SU(2)} {Yang-Mills-Higgs} theory {\em Phys.\ Rev.\ D\/} {\bf
  77} 025019

\bibitem{Kreiss1973}
Kreiss H~O and Oliger J 1973 Methods for the approximate solution of time
  dependent problems ({\em Global Atmospheric Research Programme Publication
  Series\/} no~10) (Geneva: International Council of Scientific Unions, World
  Meteorological Organization)

\bibitem{Brill1980}
Brill D~R, Cavallo J~M and Isenberg J~A 1980 {$K$}-surfaces in the
  {S}chwarzschild space-time and the construction of lattice cosmologies {\em
  J.\ Math.\ Phys.\/} {\bf 21} 2789--2796

\bibitem{Malec2003}
Malec E and \'{O}~Murchadha N 2003 Constant mean curvature slices in the
  extended {S}chwarzschild solution and the collapse of the lapse {\em Phys.\
  Rev.\ D\/} {\bf 68} 124019

\bibitem{Bizon2010}
Bizo\'n P, Rostworowski A and Zengino\u{g}lu A 2010 Saddle-point dynamics of a
  {Yang-Mills} field on the exterior {Schwarzschild} spacetime {\em Class.\
  Quantum Grav.\/} {\bf 27} 175003

\bibitem{Bartnik1988}
Bartnik R and McKinnon J 1988 Particlelike solutions of the
  {Einstein-Yang-Mills} equations {\em Phys.\ Rev.\ Lett.\/} {\bf 61} 141--144

\bibitem{Bizon1990}
Bizo\'n P 1990 Colored black holes {\em Phys.\ Rev.\ Lett.\/} {\bf 64}
  2844--2847

\end{thebibliography}

\providecommand{\newblock}{}

 %for journal submission, comment out previous two lines

\end{document}